# Amorphous intergranular films mitigate radiation damage in nanocrystalline Cu-Zr

Jennifer D. Schuler [a,b], Charlette M. Grigorian [a], Christopher M. Barr [b], Brad L. Boyce [b], Khalid Hattar [b], Timothy J. Rupert [a,*]

[a] Department of Materials Science and Engineering, University of California, Irvine, CA 92697, USA
[b] Material, Physical, and Chemical Sciences, Sandia National Laboratories, Albuquerque, New Mexico 87185, USA
[*] Corresponding Author: trupert@uci.edu

**Abstract:**

Nanocrystalline metals are promising radiation tolerant materials due to their large interfacial volume fraction, but irradiation-induced grain growth can eventually degrade any improvement in radiation tolerance. Therefore, methods to limit grain growth and simultaneously improve the radiation tolerance of nanocrystalline metals are needed. Amorphous intergranular films are unique grain boundary structures that are predicted to have improved sink efficiencies due to their increased thickness and amorphous structure, while also improving grain size stability. In this study, ball milled nanocrystalline Cu-Zr alloys are heat treated to either have only ordered grain boundaries or to contain amorphous intergranular films distributed within the grain boundary network, and are then subjected to in situ transmission electron microscopy irradiation and ex situ irradiation. Differences in defect density and grain growth due to grain boundary complexion type are then investigated. When amorphous intergranular films are incorporated within the material, fewer and smaller defect clusters are observed while grain growth is also limited, leading to nanocrystalline alloys with improved radiation tolerance.

**Keywords:** Complexion, Ion Irradiation, Grain Boundary Segregation, Grain Growth, Amorphous Intergranular Films

# 1. Introduction

Future fission and fusion nuclear reactors will push the limits of current material capabilities [1-6]. For example, hotter reactor operation temperatures that enable more efficient electricity generation are projected to exceed 1000 °C at the gas outlets of Generation IV reactor designs [1, 2, 7]. In addition to high temperature demands, longer nuclear reactor lifetimes coupled with greater safety and reliability goals necessitate the need for materials with improved radiation tolerance at these elevated temperatures [5]. Point defects formed during irradiation can aggregate into larger defect structures such as voids, dislocation loops, and stacking fault tetraheda in face centered cubic metals [8]. Radiation damage alters microstructure in complex ways and the properties of damaged microstructures are still poorly understood, especially in nanostructured metals [9].

The inclusion of surfaces and interfaces within a material has been suggested to aid in defect recovery by acting as point defect sinks and recombination sites that reduce damage accumulation leading to improved radiation tolerance [10-13]. Surfaces are generally considered to be ideal sinks that serve as perfect interstitial-vacancy recombination sites, a fact utilized by nanoporous metals that have an extremely large surface-to-volume ratio [9, 14]. For example, nanoporous Au has been shown to be an unsaturable defect sink within an ideal ligament diameter versus dose rate window where defect migration to the ligament surface occurs faster than the time between collision cascades [10, 15]. The ability of an interface, such as a grain or phase boundary, to act as a sink is understood in relation to an ideal sink through the sink efficiency, which is defined as the ratio of the flux of defects to an interface to the flux of defects to an ideal sink [16]. Interfaces in the form of phase boundaries have been observed to improve radiation tolerance in oxide dispersion strengthened (ODS) steels by serving as efficient trapping sites, which enhance defect recombination and He dispersion [17-19]. Interfaces in the form of grain boundaries are thought to serve as defect sinks by absorbing interstitials, and have been observed to facilitate defect recombination through emission of the stored interstitials to vacancies, annihilating the defect pairs [10, 20]. Han et al. [21] observed that the void denuded zone width, an indirect expression of sink efficiency, is dependent on both the grain boundary misorientation and normal plane in coarse grained Cu. Nanocrystalline metals have demonstrated excellent radiation tolerance due to their large volume fraction of grain boundaries serving as defect sinks [22-25], but grain growth during irradiation usually degrades this radiation tolerance over time at high temperature and flux due to the resultant decrease in grain boundary fraction [10, 12, 22]. In addition, a scaling breakdown has been observed where grain refinement within the nanocrystalline regime alone is not sufficient to improve radiation tolerance [26, 27]. For example, Barr et al. [26] observed no change in defect density in irradiated nanocrystalline Pt with grain sizes ranging from 100 to 20 nm. Since sink efficiency is grain boundary dependent, approaches that can drastically alter grain boundary structure and chemistry in nanocrystalline metals are desirable in order to improve radiation tolerance [28].

Grain boundaries can undergo discrete transitions in structure and composition, between different *complexion* states, in response to varying thermodynamic parameters such as temperature, composition, and grain boundary character [29, 30]. Similar to traditional bulk phase transitions, grain boundaries that have undergone complexion transitions also have significantly altered behavior. For example, Luo et al. [31] observed that Bi adsorption onto Ni grain boundaries to form bilayer complexions causes liquid-metal embrittlement compared to grain boundaries without this bilayer adsorption. Another important type of complexion is a disordered nanoscale film, also called an amorphous intergranular film (AIF). AIFs, which are created through an interfacial premelting process [32], are amorphous regions at grain boundaries that form to lower the grain boundary energy, making AIFs the thermodynamically preferred grain boundary structure at high temperatures [30, 33, 34]. AIFs and the factors that impact their formation such as temperature, composition, and structure have been investigated in numerous theoretical [29, 33], computational [35-38], and experimental studies [32, 34, 39-41]. Since AIFs are the thermodynamically preferred state, they serve to increase the thermal stability of nanocrystalline metals at elevated temperatures [34, 42], suggesting they may also combat the key issue of nanocrystalline grain growth during irradiation. In addition, the AIFs are usually heavily doped and may add a kinetic component of stabilization associated with dopant drag as the boundary moves, similar to stabilization observed in ODS steels [43, 44]. Some grain size stabilization mechanisms such as solute segregation to the grain boundary [45] and Zener pinning [46, 47] typically break down with increasing temperature [48-50]. AIFs not only stabilize the grain size, but they can do so at elevated temperatures where these structures are the preferred state. For example, a new grain size stabilization regime was found in nanocrystalline Ni-W at 1200 °C (approximately 70% of the melting temperature) due to AIF formation [42]. This new region of stability appeared after traditional solute segregation had begun to fail.

In addition to stabilizing the microstructure, AIFs are hypothesized to improve radiation tolerance. Ludy and Rupert [51] observed that AIFs act as ultra-efficient point defect sinks in molecular dynamics simulations, able to absorb both interstitials and vacancies due to their excess free volume, compared to ordered grain boundaries which preferentially accommodate interstitials. Related studies point towards excess free volume at interfaces as beneficial for radiation tolerance. Extra free volume present at phase boundary interfaces in materials such as multilayer Cu-Nb [12, 52], and Cu-V [12, 53] increases the solubility of He and minimizes blistering [54], which leads to increased He trapping and improved radiation tolerance [55-57]. Samaras et al. [58] also observed that areas of high free volume within a grain boundary can serve as defect recombination sites. The importance of free volume was reinforced experimentally by Su et al. [59], who showed that the interfaces of amorphous SiOC/crystalline Fe multilayers act as efficient point defect sinks. AIFs can also act as fast diffusion pathways, a property utilized during solid state activated sintering [33, 60], which may lead to faster in-boundary defect mobility and annihilation rates [61].

Prior research has shown that AIF formation is promoted in alloys that have dopant segregation to the grain boundary and a negative enthalpy of mixing [62]. Cu doped with Zr satisfies these criteria and nanocrystalline Cu-Zr alloys have been shown previously to form AIFs [34, 40]. In this study, the impact of AIFs on radiation tolerance is evaluated experimentally by comparing the response of nanocrystalline ball milled Cu-Zr samples heat treated to either have only ordered grain boundaries or to contain AIFs distributed within the grain boundary network. The samples were then subjected to in situ transmission electron microscopy (TEM) irradiation and ex situ irradiation. Differences in the microstructural evolution as well as total defect cluster area and number densities are evaluated from the in situ TEM irradiation data, while differences in grain growth as a function of depth and defect number density are evaluated from the ex situ irradiation data. The combination of two different irradiation modalities allows the observed trends to be compared and confirmed for the two types of materials. Both experiments confirm that there was less radiation damage and grain coarsening in the sample containing AIFs when irradiated at 25 °C. These results highlight the importance of grain boundary structural transitions on radiation tolerance, where the incorporation of AIFs distributed within the grain boundary network acts to improve the radiation damage tolerance of nanocrystalline Cu-Zr.

## 2. Materials and Methods

Cu-Zr powder with 3 at.% Zr, measured using energy dispersive X-ray spectroscopy (EDS) inside of a scanning electron microscope (SEM), was prepared using mechanical ball milling in a SPEX SamplePrep 8000M Mixer/Mill to produce particles that are hundreds of micrometers in diameter and contain nanocrystalline grains. The powders were milled for 10 h using a hardened steel vial and milling media with 2 wt.% stearic acid added as a process control agent. Powder samples were then encapsulated under vacuum in high purity quartz tubes and annealed at 950 °C (0.98$T_{solidus}$ of Cu-3 at.% Zr) for 1 h to achieve Zr dopant segregation and grain boundary structural transitions to AIFs. The samples were then either slowly cooled over a period of approximately 5 min or rapidly quenched in water in under 1 s. Slow cooling gives any AIFs formed during the high temperature treatment sufficient time to return to an ordered grain boundary structure that is preferred at lower temperatures, so this processing condition will be referred to as the *ordered grain boundary sample*. Rapidly quenching instead leads to the AIFs accessed at elevated temperatures to be frozen into place, so this specimen will be referred to as the *AIF-containing sample*. Since AIFs are only thermodynamically preferred at high temperature, it is important to remember that the AIFs quenched into place are therefore only metastable at room temperature [29, 40, 63]. It is also important to note that AIFs do not exist at every grain boundary because of the heterogeneous distributions of grain boundary energy, structure, and local chemistry [30, 40]. Past studies have shown that AIFs tend to form at grain boundaries with high relative solute excess [64]. Since high energy grain boundaries may accommodate more solute segregation [36, 65], and ball-milled nanocrystalline metals have been shown to have a higher grain boundary energy than a fully-equilibrated high angle grain boundary [66], it is expected (and has been confirmed multiple times [34, 39, 40]) that AIFs can readily form in this ball milled nanocrystalline alloy.

In situ TEM irradiation was performed during TEM inspection, while ex situ irradiation was performed on powder particle specimens under vacuum in a separate irradiation chamber. All irradiation experiments were performed using a Au$^{4+}$ ion beam generated by a 6 MV Tandem Van De Graaff accelerator at Sandia National Laboratories [67]. The projected ion beam penetration depth is the depth of peak ion concentration where the irradiating ion is eventually stopped within the material [68]. Estimations of the depth dependent damage accumulation level, projected ion beam penetration depth, and amount of implanted Au in the irradiated samples were calculated using the Stopping and Range of Ions in Matter (SRIM) 2013 program [68] with the quick Kinchin-Pease damage estimation method [69] and threshold displacement energies of 30 eV for Cu and 40 eV for Zr [27]. In situ experiments were performed on the in situ ion irradiation TEM (I$^3$TEM) JEOL 2100 at Sandia National Laboratories [67] operating at 200 kV using a 2.8 MeV Au$^{4+}$ beam at a flux of $1.5 \times 10^{11}$ ions/cm$^2$s, giving a maximum fluence of $4.0 \times 10^{14}$ ions/cm$^2$ and maximum damage level of approximately 6.8 dpa (dose rate of $2.6 \times 10^{-3}$ dpa/s). Video was collected at 15 frames per second in bright field TEM mode using a 1k × 1k camera. The in situ TEM irradiation experiments were performed continuously, without interruption. Irradiation and video collection were never paused during the entire irradiation experiment (from 0 to 6 dpa) in order to eliminate uncontrolled and unequal equilibration times between samples. The ion beam impacted the TEM samples at 60° to the sample surface in order to optimize the coincidence of the electron and ion beams. The SRIM calculations for the in situ TEM experimental setup were performed using a 60° incident angle to account for the sample tilting necessary for concurrent imaging and irradiation. Ex situ irradiations were performed using a 20 MeV Au$^{4+}$ beam at a flux of $1.1 \times 10^{12}$ ions/cm$^2$s to a fluence of $2.1 \times 10^{16}$ ions/cm$^2$ and damage level of approximately 95 dpa (dose rate of $4.8 \times 10^{-3}$ dpa/s). Irradiation occurred through rastering of the ion beam at 90° to the sample surface. The in situ irradiations were performed at 25 °C, while the ex situ irradiations were performed at nominal chamber temperatures of 25 °C and 200 °C to access different defect annihilation and mobility regimes. No active cooling was present during the in situ and ex situ irradiations, but irradiation induced heating is expected to be the same for the ordered and AIF-containing samples since they received the same flux for each experiment and have similar thermal properties. Details of the irradiation experiments are presented in Table 1.

All electron transparent samples of nanocrystalline Cu-Zr for the in situ TEM irradiations and post ex situ irradiation TEM inspections were made with the focused ion beam (FIB) lift-out technique for cross-sectional TEM inspection using an FEI Nova 600 Nanolab and FEI Quanta 3D FEG SEM, with final 5 kV polishing to remove any FIB damage. Bright field TEM, high resolution TEM, scanning TEM (STEM), STEM-EDS, and selected area electron diffraction were performed using a JEOL JEM-2100 and JEOL 2800 TEM operating at 200 kV. High resolution TEM imaging was used to study grain boundary structure using Fresnel fringes to identify interfacial films and ensure edge-on orientation of the grain boundaries. STEM-EDS maps were collected using the Cliff-Lorimer method with no absorbance with a $3 \times 3$ kernel size. Electron energy loss spectroscopy (EELS) was performed using a JEOL Grand ARM300CF in STEM mode operating at 300 kV to compute the thicknesses of the electron transparent specimens for volumetric defect area and number density calculations. The thicknesses were calculated using the log-ratio (absolute) method with a measured convergence semi-angle of 21 mrad, collection semi-angle of 30 mrad, and effective atomic number of 29. The average thicknesses of each electron transparent sample were measured by finding the average sample thickness from a total measured area of approximately 12 $\mu m^2$ on each TEM foil.

Transmission Kikuchi diffraction (TKD) was performed on the ex situ irradiated samples using an FEI Quanta 3D FEG SEM-FIB with Kikuchi diffraction patterns collected using an Oxford Nordlys F+ electron backscatter diffraction detector. The electron transparent samples were held at a working distance of approximately 3.5 mm and tilted 20° from horizontal, while the SEM was operated at 30 kV and 11 nA using a 1 mm aperture with the microscope in its high current analytical mode [70-72]. TKD maps were collected using a step size of 10 nm in a series of small maps in order to keep scan times short to best minimize drift. Multiple scans were then combined to create one large map for each sample, where each combined map contained 175,000 to 250,000 points. A minimum grain reconstruction threshold of 2° was used in TKD data analysis. No other data processing was performed, so the TKD results shown in this study are comprised of the raw data. Grain size was calculated from at least 70 grains clearly identified in bright field TEM micrographs by measuring the areas of grains and calculating the average circular equivalent diameter.

## 3. Results

### 3.1. Initial Microstructure and Radiation Conditions

Figure 1 shows representative TEM images of the Cu-Zr samples after receiving heat treatments to create the ordered grain boundary and AIF-containing samples. The average grain size before irradiation was 59 ± 23 nm for the ordered grain boundary sample and 54 ± 20 nm for the AIF-containing sample. Figure 1(a) shows a bright field TEM image of the AIF-containing sample microstructure before irradiation. Figure 1(b) shows a high resolution TEM image of an AIF from the AIF-containing sample, where the dashed red lines highlight where the AIF meets the two bounding crystals. Figure 1(c) shows a dark field STEM image from the AIF-containing sample and Figure 1(d) shows the associated STEM-EDS map from this same area with the distribution of Zr within the microstructure shown in yellow. Zr segregation to the grain boundaries was present in both samples, with two grain boundaries indicated by red arrows in these figures. Due to the ball milling process, a small amount of ZrC precipitates are also found within the microstructure [34], estimated as approximately 1 vol.% ZrC with average sizes of 20 nm and 5 nm when found at the grain boundary and in the grain interior, respectively, according to prior measurements on these samples from Grigorian and Rupert [39]. The inset in Figure 1(a) shows the associated selected area electron diffraction pattern, where the solid blue lines indicate the face centered cubic Cu phase and the dashed red lines indicate the ZrC phase. The presence of ZrC precipitates could impact the radiation damage tolerance of the nanocrystalline Cu-Zr samples since precipitates can both limit grain boundary mobility through pinning [46, 47] and serve as defect sinks [18, 73]. However, both the ordered grain boundary and AIF-containing samples were sourced from the same ball milled powder and contained the same size and density of ZrC precipitates. Therefore, any differences in radiation damage tolerance between the two samples can be attributed solely to the presence or absence of AIFs in the grain boundary network.

Figure 2 shows the estimated damage accumulation level as a function of depth into the sample for each experiment. Figure 2(a) shows damage level (blue circles) of the in situ TEM irradiation, where the dashed black line at 95 nm indicates the average thickness of the TEM specimens used for these experiments. The associated gray box indicates the approximate thickness range subjected to irradiation after accounting for a 15 nm standard deviation of the TEM sample thickness. SRIM calculations estimate a maximum possible value of ~0.03 at.% Au implanted into the TEM foil. Figure 2(b) shows a similar calculation for the damage level (blue circles) and amount of implanted Au (red circles) for the ex situ irradiation experiment. SRIM calculations estimate a maximum damage level of approximately 95 dpa at a depth of ~2 μm within the sample, and a maximum of ~0.3 at.% Au implanted. While it is possible that Au implantation could alter local alloy chemistry and possibly impact behavior upon irradiation, only a small amount of Au was implanted and both the ordered grain boundary and AIF-containing samples were subjected to the same degree of Au implantation during the in situ TEM and ex situ irradiations. Again, any differences in radiation damage tolerance can be attributed solely to differences in the grain boundary structure.

The sink efficiencies of different grain boundaries have previously been indirectly measured using denuded zone widths, where a larger denuded zone correlates to improved sink efficiency. However, several factors limit this interpretation [74]. For example, the grain boundary structure itself can be dynamic during irradiation [61]. Also, AIFs do not occur at every grain boundary, further restricting the comparison of individual grain boundaries for sink efficiency and implications on radiation tolerance. Instead the grain boundary network is investigated as a whole for its impact on radiation tolerance through analysis of grain growth, as well as defect area and number densities within the different samples.

## 3.2. Microstructural Evolution and Defect Cluster Area and Number Densities from In Situ TEM Irradiation

Microstructural evolution during in situ TEM irradiation was analyzed using bright field TEM images gathered from the in situ TEM irradiation videos. While the exact values for the measured defect densities can be influenced by experimental variables such as viewing magnification, both the ordered grain boundary and AIF-containing samples were analyzed with the same methods. In addition, any trends extracted from our in situ experiments will be confirmed with the ex situ studies. Significant contrast changes were observed to occur in each sample during the course of irradiation. These contrast changes can be caused by grain growth, change in grain shape, defect formation, and defect accumulation, as well as TEM driven artifacts such as bend contours. While both the ordered and AIF containing samples had contrast shifts, they manifested in physically different ways in each sample. Two types of microstructural evolution were observed during the in situ TEM irradiation experiments. Type 1 was a general change in grain shape without a significant change in area, which may be due to grain boundary motion or relaxation of grain boundary structure. Type 2 is grain growth, indicated by a significant increase in grain area at the expense of a neighboring grain. Figure 3 presents bright field TEM images captured from the in situ irradiation videos for the ordered grain boundary and AIF-containing samples. Figures 3(a)-(c) show the microstructural evolution of the ordered grain boundary sample with increasing dose at 1, 3, and 6 dpa, respectively. Figures 3(d)-(f) shows similar images for the AIF-containing sample. Grains indicative of each microstructural evolution type are highlighted in red and numbered in accordance with their evolution mode in Figure 3. Examples of both types of microstructural evolution were observed in the ordered grain boundary sample, but Type 2 grain growth was not observed in the AIF-containing sample within the imaged region. Grain growth observed at grain 2b in the ordered grain boundary sample from Figures 3(a)-(c) is shown in greater detail in Figure 4. The red arrow in Figure 4(d) indicates a region of significant contrast change in which the grain is then observed to grow. Figures 4(d)-(i) from 4.0 to 4.5 dpa are shown in 0.1 dpa increments in order to provide greater detail during the dose period in which the grain undergoes the most growth. In the ordered grain boundary sample, irradiation defects, grain shape change, and several instances of grain growth were observed. In contrast, the AIF-containing sample contained examples of irradiation defects and grain shape change, but no discernible grain growth. We acknowledge that it is difficult to make a strong statement about changes (or lack thereof) to quantities such as average grain size when only a relatively small collection of grains can be imaged during the continuous in situ TEM irradiation experiment. Therefore, we will return to this issue during our discussion of the ex situ irradiated samples.

In addition to different physical manifestations of the contrast changes shown in Figure 3 and Figure 4, the ordered grain boundary and AIF-containing samples had different degrees or amounts of contrast change. The degree of contrast change between the samples can be shown by generating a difference map of contrast shifts between images taken at the beginning and end of the irradiation experiment (i.e., at 0 and 6 dpa, respectively). This was done by first adjusting the 0 and 6 dpa bright field TEM images to have the same brightness and contrast scaling. The images were aligned using StackReg in ImageJ [75], and then the 6 dpa image was subtracted from the 0 dpa image using the Image Calculator function. The bright field TEM brightness and contrast adjusted images used for subtraction are shown in Figures 5(a) and (b) for the ordered grain boundary sample, and Figures 5(d) and (e) for the AIF-containing sample. The difference maps for each sample are then shown in Figures 5(c) and (f), where grayscale pixel values have been mapped to a color gradient scale for easier inspection. Negative values refer to a darker grayscale shift in the bright field TEM images, and positive values refer to a lighter grayscale shift. Larger contrast shifts appear as orange and white, whereas smaller contrast shifts appear as black and purple. The increased amount of contrast shift in the ordered grain boundary sample compared to the AIF-containing sample provides an additional measure of increased microstructural evolution. These contrast changes can be due to all of the mechanisms identified above, including defect formation, defect accumulation, grain rotation, grain shape change, or grain growth. Since TEM contrast artifacts such as bend contours will also register as a difference in Figure 5, ex situ data is necessary to complement these in situ TEM results.

Next, we investigated the population of irradiation-induced defects in each sample. Determination of defect density, calculated using defect width, area, number, or some combination thereof, has four notable experimental limitations [76]: (1) The resolution limit of the imaging technique used curtails measurement of the smallest defects, (2) defects can be lost due to diffusion to the TEM foil surface, (3) differences in foil thickness influence perceived defect densities, and (4) only a subset of defects are resolvable due to imaging condition constraints. The first three limitations are taken into consideration by using identical TEM inspection conditions between samples and by measuring the sample thickness using EELS to normalize the defect densities by the inspected volume. The fourth limitation poses the greatest challenge. Performing traditional TEM techniques to determine radiation defect nature (e.g., as described by Jenkins and Kirk [76]) involves tilting the TEM sample to precise imaging conditions so that the appropriate coupling of the diffraction vector ($g$) and the Burgers vector ($b$) are met to yield $g \cdot b$ visibility or invisibility conditions for defect analysis, but this is challenging in nanocrystalline metals due to the small grain size [28]. While the exact nature of the defects are not identified here, previous heavy ion irradiation studies using Cu observed that vacancy and interstitial clusters form directly within the collision cascade and then coalesce into point defect clusters, loops, dislocations, dislocation tangles, cavities, and vacancy-type stacking fault tetrahedra [77, 78]. Irradiation induced defects were isolated compared to other TEM contrast features such as bend contours through feature movement using frame by frame video analysis. Irradiation induced defects were identified as discrete dark contrast features that appeared and evolved asynchronously within a grain, moving in a jerky manner consistent with defect percolation motion [79, 80], compared to bend contours that move smoothly and synchronously [81]. Defects were identified by analyzing the video frame-by-frame in order to isolate those features within the grain that exhibited dark contrast and moved in a jerky, non-continuous manner, and were then processed and analyzed using GIMP and ImageJ. All video images used for in situ TEM defect analysis were adjusted to have the same brightness and contrast levels, and then defects were manually identified and cross-referenced with images from other dose values in order to determine the defect size and shape. Most importantly, this common procedure was consistently applied across all images used to obtain the data for Figure 6, which allows trends between the samples to be compared.

Figure 6 presents the total defect cluster area normalized by volume, hereafter referred to as the *defect area density*, measured using the in situ TEM irradiation video from 0 to 6 dpa. The scale bars in Figures 6(a)-(d) are all 100 nm. Figures 6(a) and (c) show bright field TEM images of an example grain from the ordered grain boundary sample subjected to in situ TEM irradiation at 1 and 2 dpa, with the grain outlined with a dashed red border. The radiation-induced defects present in each grain are marked in black in Figures 6(b) and (d). All black spot features inside the selected grain in Figures 6(a)-(d) are those that moved in a non-continuous manner. They may be new defects created by the ion irradiation or they may be pre-existing defects from the ball milling process used to prepare the powder sample that then move due to irradiation. The defect area density is then calculated by finding the total area of identified defects and then dividing by the grain volume, which was calculated using the sample thickness multiplied by the respective outlined grain area, with an example shown in Figures 6(a) and (c). The thickness of the electron transparent specimens was $86 \pm 11$ nm for the ordered grain boundary sample and $103 \pm 10$ nm for the AIF-containing sample. Figure 6(e) shows the same grain from Figures 6(a) and (c), but rotated to schematically show the impact of specimen thickness on the measurement of defect area density. Since bright field TEM images show features in projection, the true shape and size of the defect may be different than what is viewed in a bright field TEM image. For example, the solid line represents the true defect shape, while the dotted line represents the perceived defect shape in the projected bright field TEM image. A thicker sample would not only make the dislocation lines appear longer, but more defects would be captured in projection creating a higher perceived defect density if changes to the sample volume were ignored. Figure 6(f) shows defect area density with increasing dose for the ordered grain boundary (blue circles) and AIF-containing (red squares) samples, measured from 10 grains in each sample. Error bars are calculated by adjusting the measurement volume due to variation in measured sample thickness. In other words, the lower bound is the total defect cluster area divided by the maximum measurement volume while the upper bound is calculated using the minimum measurement volume. Both samples show an appreciable defect number density at 0 dpa which indicates there is population of pre-existing defects within the sample before irradiation. These pre-existing defects can also move and evolve due to irradiation. Defects present at 0 dpa (as-prepared) may be defects stored during the processing of the material or due to the FIB sample preparation method. Both samples begin the in situ irradiation experiment with similar defect area densities (0.30 $\mu m^2/\mu m^3$ for the ordered grain boundary sample and 0.27 $\mu m^2/\mu m^3$ for the AIF-containing sample), but differences become more noticeable as dose increases. The ordered grain boundary sample shows a higher defect area density compared to the AIF-containing sample. At 6 dpa, the ordered grain boundary sample had approximately 1.5 times the defect area density as the AIF-containing sample. From 2 to 6 dpa, both samples also have a positive slope (linear fits of this region give slopes of ~0.18 $\mu m^2/\mu m^3 dpa$ for the ordered grain boundary sample and ~0.05 $\mu m^2/\mu m^3 dpa$ for the AIF-containing sample), but p-value statistical analysis indicates that the difference between the two slopes is statistically significant ($\alpha = 0.05$). This means that while both samples are not yet in a state of defect saturation, the AIF-containing sample is accumulating defects at a slower rate.

In addition to defect area density, the number of defects formed (hereafter referred to as *defect number density*) was also measured from the in situ TEM irradiation experiments. Figure 7(a) shows the defect number density for the ordered grain boundary (blue circles) and AIF-containing (red squares) samples with increasing dose. Error bars again represent the upper and lower bounds of defect number density due to variations in sample thickness. Dividing the values in Figure 6(f) by the values in Figure 7(a) allows an estimate of the area of an average defect (hereafter referred to as the *average defect area*) to be calculated. Figure 7(b) shows the average defect area for the ordered grain boundary (blue circles) and AIF-containing (red squares) samples. Figures 7(a) and (b) show that the defect number density between the two samples is similar but the average defect area diverges, with the ordered grain boundary sample having a larger average defect area than the AIF-containing sample after ~2 dpa. The larger average defect area indicates more defect coalescence in the ordered grain boundary sample than in the AIF-containing sample. A high density of grain boundaries (i.e., a smaller grain size) interrupts the mean free path of migrating defects and can also increase defect sinking [82]. Prior atomistic simulations have shown that AIFs can act as ultra-efficient defect sinks [51], which will cause even more defect sinking compared to a sample with only ordered grain boundaries. These factors will cause fewer defects to be available in the grain interior for defect coalescence in the AIF-containing sample compared to the ordered grain boundary sample [82].

The normalized defect cluster area, defect number density, and average defect area values presented in Figures 6 and 7 were gathered from frames collected from the in situ TEM irradiation video. We remind the reader that the images were taken during a continuous irradiation experiment, which was not stopped in order to collect images. The data at 0 dpa represents the as-prepared material state, with irradiation occurring until each sample reached the 6 dpa state.

Therefore, the images used to generate Figures 6 and 7 represent the instantaneous state of the irradiated sample and not an equilibrated collection of defects. We also note that the defect measurements in Figure 6 and 7 are measurements of the defect population inside of the grain structure, but do not necessarily provide insight into the behavior of any individual defect. For example, the relatively constant values for the average defect size for the AIF-containing sample in Figure 7(a) does not mean that all defects are stagnant and nothing in the microstructure changes. Individual defects still grow, shrink, or disappear, but the average defect size in the AIF-containing sample remains roughly constant.

**3.3. Grain Growth and Defect Number Density from the Ex Situ Irradiations at 25 °C and 200 °C**

While in situ TEM irradiation provides crucial insight regarding damage mechanisms and time dependent phenomena, interpretation can be affected by factors such as surface defect annihilation [83, 84] and electron beam effects [85-87]. Complementary ex situ irradiations were performed using the conditions presented in Table 1 and Figure 2(b). TKD measurements from these samples are shown in Figures 8 and 9. In the ex situ sample reference frame, the ion beam impacts normal to the sample surface along the Y-axis and the cross-sectional TEM inspection is along the Z-axis. The Z-axis was chosen for inverse pole figure (IPF) grain orientation analysis since differing textures between samples along the inspection axis may impact defect density comparisons, and is referred to as IPF-Z. Equal area projection plots are used to analyze the IPF-Z maps, where an even distribution of plotted directions across the projection area correlates to a random distribution of crystal directions in the measurement region and a lack of texture [88]. Irradiation was performed at 25 °C and 200 °C in order to alter the defect mobility regime being accessed. Irradiation in the medium temperature regime where both interstitials and vacancies are mobile occurs between Stage III and Stage V recovery. In Cu, Stage III recovery where vacancies become mobile begins at approximately -3 °C and Stage V recovery defined by vacancy-impurity migration begins at approximately 277 °C [89]. Therefore, ex situ irradiation at 200 °C probes closer to Stage V recovery regime whereas ex situ irradiation at 25 °C probes closer to Stage III. It is important to note that Cu-Zr may have different Stage III and Stage V recovery onset temperatures compared to pure Cu due to its slightly lower melting temperature, dopant chemistry, and nanocrystalline microstructure [89]. AIFs may also destabilize at these temperatures and revert back to ordered grain boundaries, since it is less than 60-85% of the material's melting temperature, where AIFs are thermodynamically preferred [34, 36, 63, 90]. Experimental limitations prevented hotter irradiation due to the adhesive necessary to properly mount the powder sample for irradiation. Swelling is another important irradiation damage behavior that peaks in Cu at approximately 300 °C [89]. Future work is needed in order to evaluate the effect of complexion type on the swelling behavior of irradiated nanocrystalline alloys.

Figures 8(a)-(c) show the bright field TEM, Kikuchi band contrast map, and IPF-Z maps respectively for the ordered grain boundary sample irradiated at 25 °C (outlined in purple). Figures 8(g)-(i) show similar images for the AIF-containing sample (outlined in orange). On each image, the dashed red lines denote the sample surface and the dashed yellow lines denote the projected ion beam penetration depth, ~2.2 μm deep within the sample. Figure 8(e) shows the sample reference frame axes while Figures 8(d) and (f) show equal area projection plots for the corresponding IPF-Z maps. The IPF-Z equal area projection plots were only collected for the regions most affected by irradiation from the sample surface to the projected ion beam penetration depth (i.e., between the red and yellow dashed lines). Both samples irradiated at 25 °C have similar textures within the projected ion beam penetration depth, with a slight preference for (101) oriented grains. Figure 9 shows similar data to that presented in Figure 8, but for the ex situ irradiation

experiments at nominally 200 °C. The equal area projection distribution plots in Figures 9(d) and (f) again show a preference for the (101) grain orientation.

Figure 10 shows the grain size as a function of cross-sectional depth measured from the irradiated samples in Figures 8 and 9. Each point shows the average grain size that was measured within a ~100 nm-thick segment beneath the irradiated surface. For example, the first point shows the average size of the grains present from the sample surface at 0 nm to 100 nm deep, while the second point shows the average size of those grains that are 100 nm to 200 nm deep within the sample. At least 10 grains were used to find the average grain size in each segment. While one would prefer to use hundreds of grains to calculate an average grain size at each depth, only a limited number of grains are present and in good diffraction conditions at a given depth within the TEM sample, necessitating the usage of few measurements. The data should be considered in a spatial context with its neighboring data and compared against the initial average grain size (grey dashed line) and the standard deviation of the initial grain size (represented by the shaded box around the average) that were calculated from ~100 grains. The grain position beneath the irradiated surface was determined using the centroid of the grain, where the centroid is defined as the average coordinates of all of the image pixels contained within the grain. The vertical dashed gray line shows the region of maximum dose coinciding with the projected ion beam penetration depth at ~2.2 μm. Figure 10(a) shows the grain size measurements for the samples irradiated at 25 °C, where the ordered grain boundary sample (blue circles) demonstrates clear grain growth to a grain size of ~85 nm compared to the initial average grain size of 59 nm. In the portions of the cross-section deeper than the maximum dose, the average grain size returns to the expected baseline grain size for the ordered grain boundary sample. The AIF-containing sample (red triangles) demonstrates a different trend, where within the irradiated region the average grain size stays close to the baseline grain size of 54 nm. There is only even the possibility of some slight grain growth at the maximum dose region, where damage levels reach 95 dpa and maximum Au implantation occurs that could alter the expected alloy chemistry. Figure 10(b) shows similar grain size data for the samples irradiated at 200 °C. Both the ordered grain boundary (green circles) and AIF-containing (purple triangles) samples demonstrate nearly identical behaviors, with an increase in grain size to approximately 100 nm and then a return to their respective baseline grain sizes beyond the maximum dose region.

In addition to microstructural changes, irradiation defect clusters were also investigated. Similar to challenges faced in the in situ TEM defect analysis, defect counting is dependent on magnification and imaging conditions, but both the AIF-containing and ordered samples were counted with the same methods, allowing for a comparison of trends. Figure 11 presents TEM images of different types of defects observed in the irradiated samples. Figure 11(a) shows an assortment of defects observed in all of the ex situ irradiated samples present as black spot damage which could be point defect clusters, stacking fault tetrahedra, loops, dislocations, or dislocation tangles [77, 78]. A possible stacking fault tetrahedron from the 25 °C irradiated ordered grain boundary sample, indicated by the red arrow in Figure 11(a), is shown in greater detail in Figure 11(b), where an inverse fast Fourier transform is used to filter the image. Dotted red lines frame the possible stacking fault tetrahedron and solid red lines indicate a potential stacking fault inside the tetrahedron. Further TEM inspection is necessary to confirm the exact nature of the defect [76], as such contrast may also be caused by other crystallographic defects such as dislocation cross-slip or a dislocation lock. Cavities can be easily viewed by slightly defocusing in the TEM, with cavities appearing bright with a dark border in an under-focused condition [76]. Figure 11(c) shows several cavities from the 200 °C ex situ irradiation AIF-containing sample with one cavity indicated by a red arrow and shown in greater detail in Figure 11(d).

The defect number density for the ex situ irradiated samples was calculated by counting the number of black spot damage features visible in bright field TEM images within the projected ion beam penetration depth (sample surface to 2.2 μm deep) and then divided by the sample measurement volume. The sample measurement volume is the combined grain area from those grains in which defects were visible multiplied by the sample thickness. Error bars are calculated using the same method as in Figure 6(f) and Figure 7(a) by adjusting the measurement volume due to variation in sample thickness. The ex situ irradiation defect number density data was collected from images gathered long after irradiation was complete, and can be viewed as being representative of the equilibrated state. Figure 12 shows the defect number density for the ordered grain boundary and AIF-containing samples irradiated at 25 °C and 200 °C. For the samples irradiated at 25 °C, the ordered grain boundary sample is shown with a blue circle and the AIF-containing sample with a red circle. For the 200 °C irradiation, the ordered grain boundary sample is shown with a green square and the AIF-containing sample with a purple square. For the 25 °C irradiation, the ordered grain boundary sample showed a higher defect number density compared to the AIF-containing sample, well beyond any considerations associated with variations in sample thickness, leading to the conclusion that there was improved defect sinking in the AIF-containing sample. For the 200 °C irradiation, both the ordered grain boundary and AIF-containing samples had comparable defect number densities.

The defect number densities for the in situ TEM irradiation samples in Figure 7(a) show no discernible difference between the ordered grain boundary and AIF containing samples, whereas the defect number densities for the ex situ irradiation samples in Figure 12 show a substantial difference in the 25 °C irradiation condition. This may be caused by increased defect sinking during in situ TEM irradiation due to the presence of nearby free surfaces. Kirk et al. [91] have used a comparison of in situ TEM ion irradiation and bulk neutron irradiation of Mo to show that dislocation loop densities can become comparable, although this finding is dependent on the dose and dose rate. Therefore, there may exist a set of irradiation conditions for nanocrystalline Cu-Zr where the defect losses due to nearby free surfaces are negated by the dose and dose rate, making in situ TEM ion irradiation results directly comparable to ex situ irradiation results. Future work is needed to determine if the irradiation conditions used in this study are within this range. There is also significantly more grain growth during the harsher ex situ irradiation that decreases the probability that a defect will make it to the grain boundary network. This grain growth can be attributed to the higher fluence achieved during the ex situ irradiations, which provides more thermal energy and may also impact defect density. The defect number densities of the in situ TEM irradiation samples in Figure 7(a) are also ~25 times higher than that of the ex situ irradiation samples in Figure 12, despite using a substantially lower dose. In addition to the possible thermal annealing of defects, this may also be due to the higher magnification at which defects were measured and counted for the in situ TEM irradiation samples compared to the lower magnification used for the ex situ irradiation samples. This allowed a higher number of small defects to be observed for the in situ TEM irradiation samples, details which were likely missed during the examination of ex situ samples or destroyed by sample preparation to make an electron transparent foil, causing higher defect number densities to be measured from the in situ irradiation. However, we emphasize that consistent imaging conditions and magnifications were used within a given set of experiments (in situ or ex situ irradiation), meaning that comparison of the relative values between ordered grain boundary and AIF-containing specimens are reliable.

## 4. Discussion

For irradiations at 25 °C, the AIF-containing samples have a lower defect area density (in situ), smaller average defect area (in situ), and lower defect number density (ex situ) compared to the ordered grain boundary samples. Hence, all signs point to the AIF-containing sample having improved radiation tolerance at 25 °C compared to its ordered grain boundary counterpart. Prior molecular dynamics simulations have shown that AIFs act as unbiased sinks due to their increased thickness and excess free volume compared to ordered grain boundaries [51]. The increased thickness of an AIF also allows more of the collision cascade itself to be contained within the boundary compared to an ordered grain boundary, allowing defects to be produced closer to the sink and enabling faster recombination rates [51]. AIFs may also have different defect production rates compared to an ordered grain boundary due to the presence of an amorphous zone instead of being fully crystalline. Moreover, AIFs can act as fast diffusion pathways [33, 60, 61], which may further increase in-boundary diffusion and recombination rates. AIFs may also improve radiation tolerance through easy grain rotation. Texture from grain rotation has been observed to occur upon irradiation to create easy ion channeling paths [92]. Shear transformation zones present at AIFs may further allow grain rotation into the easiest channeling alignment without grain growth [93]. Slight textural changes between the ordered grain boundary and AIF-containing samples in the irradiated regions shown in Figure 8 could indicate this damage accommodation mechanism without grain growth. As a result, incorporation of AIFs can reduce the overall defect density and improve radiation tolerance.

The 25 °C irradiated AIF-containing samples also have less irradiation-induced grain growth compared to the ordered grain boundary samples. Grain growth during irradiation can be caused by a number of factors including ion beam heating, radiation induced diffusion, dopant grain boundary desegregation, and ion beam mixing. Radiation induced diffusion is caused by atomic jumps within thermal spikes that generate the movement of atoms across grain boundaries. This in turn causes grain growth that is dependent on the local grain boundary curvature, nature of the cascade structure, formation of subcascades at grain boundaries [94-97], intrinsic properties of the material, and radiation conditions [28]. The increased thickness of an AIF compared to an ordered grain boundary may capture more of these atomic jumps and impede atomic movement into neighboring grains, thus limiting grain growth. The AIFs may also restrict the local grain boundary mobility at thermal spikes [95] since these complexions are preferred at elevated temperatures. The portion of the grain boundary network that transforms to an AIF both lowers the net grain boundary network energy and locally restricts grain boundary mobility. Prior studies of grain growth have shown that pinning only a fraction of grain boundaries can impede grain growth of the entire grain boundary network [98, 99]. Therefore by distributing AIFs within the grain boundary network the net grain boundary network energy and mobility is decreased which further prevents grain growth of the nanocrystalline alloy.

Desegregation of dopants at the grain boundary could also degrade thermodynamic grain size stabilization and cause grain growth [100-102]. For example, Cr depletion and Ni enrichment have been observed at the grain boundaries of irradiated 304L and 316L austenitic stainless steels [103], while dopant desegregation due to increasing temperature has been observed in nanocrystalline Ag-W [49]. Similarly, ion beam mixing disrupts the compositional separation between grain boundary and grain interior [104]. Miscible alloys have been observed to undergo ballistic intermixing at phase boundaries, which has been observed in multilayers of Al-Nb [105] and Fe-W [106]. In contrast, immiscible alloy multilayers such as Cu-Nb and Cu-V [12, 52] retain sharp chemical boundaries with no intermetallic formation or amorphization detected upon irradiation due to their large positive enthalpy of mixing, which indicates that any ballistic mixing is counteracted by dynamic demixing. Irradiation of amorphous SiOC/crystalline Fe multilayers, which mimic the amorphous/crystalline regions at the grain boundary when AIFs are present, showed intermixing -223.15 °C but then transitioned to demixing at 299.85 °C, indicating improved radiation tolerance at elevated temperatures [107]. Cu-Zr is a miscible alloy, chosen in part specifically for this characteristic to promote AIF formation at the grain boundaries [62]. STEM-EDS was performed on the ordered grain boundary and AIF-containing samples after both the 25 °C and 200 °C ex situ irradiation. Zr segregation to the grain boundaries was still present after irradiation, but subtle changes in adsorbed Zr and complexion transitions within the grain boundary network could possibly degrade radiation tolerance over longer times than those studied here. Even for the ordered grain boundary samples, it is worth noting that the thermal input alone is not sufficient to cause any of the observed grain growth. Annealing studies of the same ball-milled nanocrystalline Cu-Zr alloy showed excellent thermal stability at 750 °C even when ordered grain boundaries still dominated the grain boundary network [34] due to the combined effects of Zr dopant segregation that lowers the grain boundary energy and kinetic pinning from ZrC particles. It is also important to note that the ZrC particles and AIF thicknesses may also have been altered by irradiation which would in turn impact subsequent radiation tolerance. Therefore, phenomena specific to radiation such as radiation induced diffusion, dopant grain boundary desegregation, and ion beam mixing are necessary to cause the grain growth observed in the ordered grain boundary sample.

The ability of AIFs to improve radiation tolerance appears to break down in the 200 °C ex situ irradiation experiments, where the AIF-containing sample behaved similarly to the ordered grain boundary sample. Both samples had a similar defect number density and degree of grain growth within the projected ion beam penetration depth. This breakdown can be attributed to the metastable nature of an AIF that was quenched into place from a high temperature. AIFs are only the thermodynamically-preferred complexion state at temperatures of above ~60-85% of the material's melting temperature [34, 36, 63, 90]. At 200 °C, the preferred grain boundary state is ordered, meaning that an AIF quenched into place is only metastable and wants to transform back. The combined effects of the ion irradiation, irradiation induced heating, and 200 °C stage temperature were sufficient to transform enough AIFs out of their metastable state, causing them to crystallize and form ordered grain boundaries once again. It is important to note that some AIFs could still remain distributed within the grain boundary network in the AIF-containing sample irradiated at 200 °C, but there are apparently not enough to affect the overall behavior. AIFs are known to form at high temperatures close to the melting point, but irradiating within the temperature regimes where AIFs are thermodynamically preferred was not explored directly here due to limitations on the available equipment for irradiation. Future work is needed to investigate the radiation tolerance of alloys containing AIFs in the temperature regime where AIFs are thermodynamically preferred.

## 4. Summary and Conclusions

Three important conclusions can be drawn about the importance of complexion structure for radiation tolerance from experiments performed in this study by comparing the trends observed from the in situ TEM and ex situ irradiation studies. First, AIFs decrease the number and size of irradiation defects. This is evidenced by the AIF-containing samples having a lower defect area density and average defect area measured from the in situ TEM irradiation video, and a lower defect number density measured from 25 °C ex situ irradiation compared to the ordered grain boundary samples. Second, AIFs limit grain growth during irradiation. This is evidenced by the absence of grain growth in the AIF-containing sample during in situ TEM irradiation, and the 25 °C ex situ irradiation experiment. Third, complexion transitions due to radiation damage and temperature can change the radiation tolerance. This is evidenced by the divergence in behaviors for the AIF-containing samples for the ex situ irradiation experiments performed at 200 °C and 25 °C. Enhanced radiation tolerance in nanocrystalline Cu-Zr due to AIFs distributed within the grain boundary network was observed during 25 °C irradiation, but this effect was not seen after irradiation experiments at 200 °C.

A main challenge for nanocrystalline metals in nuclear applications is a loss of radiation tolerance due to grain growth, which is further exacerbated by high temperature operating conditions. In nanocrystalline Cu-Zr, the transformation of an ordered grain boundary network to an AIF-containing grain boundary network improved the overall radiation tolerance. Further experiments are needed to probe whether AIFs remain in the material during irradiation at the elevated temperatures at which they form. The grain boundary network as a whole can be used to improve the radiation tolerance of nanocrystalline alloys through a transition to an amorphous complexion structure.

**Acknowledgements**
J.D.S. and T.J.R. acknowledge support from the U.S. Department of Energy, Office of Basic Energy Sciences, Materials Science and Engineering Division under Award No. DE-SC0014232, and the U.S. Department of Energy, Office of Science, Office of Workforce Development for Teachers and Scientists, Office of Science Graduate Student Research (SCGSR) program. C.M.B., B.L.B., and K.H. were fully supported by the Division of Materials Science and Engineering,

Office of Basic Energy Sciences, U.S. Department of Energy. This work was performed, in part, at the Center for Integrated Nanotechnologies, an Office of Science User Facility operated for the U.S. Department of Energy (DOE) Office of Science. Sandia National Laboratories is a multimission laboratory managed and operated by National Technology & Engineering Solutions of Sandia, LLC, a wholly owned subsidiary of Honeywell International, Inc., for the U.S. DOE's National Nuclear Security Administration under contract DE-NA-0003525. The views expressed in the article do not necessarily represent the views of the U.S. DOE or the United States Government. The authors would like to thank Mr. D. Buller and Dr. B.R. Muntifering for their assistance.

# References

[1] T. Abram, S. Ion, Generation-IV nuclear power: a review of the state of the science, Energy Policy 36(12) (2008) 4323-4330.
[2] L.K. Mansur, A.F. Rowcliffe, R.K. Nanstad, S.J. Zinkle, W.R. Corwin, R.E. Stoller, Materials needs for fusion, Generation IV fission reactors and spallation neutron sources–similarities and differences, J. Nucl. Mater. 329 (2004) 166-172.
[3] S.J. Zinkle, G.S. Was, Materials challenges in nuclear energy, Acta Mater. 61(3) (2013) 735-758.
[4] K.L. Murty, I. Charit, Structural materials for Gen-IV nuclear reactors: challenges and opportunities, J. Nucl. Mater. 383(1-2) (2008) 189-195.
[5] R.W. Grimes, R.J.M. Konings, L. Edwards, Greater tolerance for nuclear materials, Nature Mat. 7 (2008) 683.
[6] E.E. Bloom, S.J. Zinkle, F.W. Wiffen, Materials to deliver the promise of fusion power–progress and challenges, J. Nucl. Mater. 329 (2004) 12-19.
[7] D. Chapin, S. Kiffer, J. Nestell, The Very High Temperature Reactor: A Technical Summary, MPR Associates, Alexandria, VA, 2004.
[8] G.S. Was, Fundamentals of Radiation Materials Science: Metals and Alloys, Springer, 2016.
[9] X. Zhang, K. Hattar, Y. Chen, L. Shao, J. Li, C. Sun, K. Yu, N. Li, M.L. Taheri, H. Wang, Radiation damage in nanostructured materials, Prog. Mater Sci. 96 (2018) 217-321.
[10] I.J. Beyerlein, A. Caro, M.J. Demkowicz, N.A. Mara, A. Misra, B.P. Uberuaga, Radiation damage tolerant nanomaterials, Mater. Today 16(11) (2013) 443-449.
[11] W. Han, M.J. Demkowicz, N.A. Mara, E. Fu, S. Sinha, A.D. Rollett, Y. Wang, J.S. Carpenter, I.J. Beyerlein, A. Misra, Design of radiation tolerant materials via interface engineering, Adv. Mater. 25(48) (2013) 6975-6979.
[12] X. Zhang, E. Fu, A. Misra, M.J. Demkowicz, Interface-enabled defect reduction in He ion irradiated metallic multilayers, JOM 62(12) (2010) 75-78.
[13] A. Misra, M.J. Demkowicz, X. Zhang, R.G. Hoagland, The radiation damage tolerance of ultra-high strength nanolayered composites, JOM 59(9) (2007) 62-65.
[14] N.J. Briot, M. Kosmidou, R. Dingreville, K. Hattar, T.J. Balk, In situ TEM investigation of self-ion irradiation of nanoporous gold, J. Mater. Sci. 54(9) (2019) 7271-7287.
[15] E.M. Bringa, J.D. Monk, A. Caro, A. Misra, L. Zepeda-Ruiz, M. Duchaineau, F. Abraham, M. Nastasi, S.T. Picraux, Y.Q. Wang, D. Farkas, Are nanoporous materials radiation resistant?, Nano Lett. 12(7) (2012) 3351-3355.
[16] A.P. Sutton, R.W. Balluffi, Interfaces in Crystalline Materials, Oxford University Press, New York, 1995.
[17] L.L. Hsiung, M.J. Fluss, S.J. Tumey, B.W. Choi, Y. Serruys, F. Willaime, A. Kimura, Formation mechanism and the role of nanoparticles in Fe-Cr ODS steels developed for radiation tolerance, Phys. Rev. B 82(18) (2010) 184103.
[18] C. Lu, Z. Lu, X. Wang, R. Xie, Z. Li, M. Higgins, C. Liu, F. Gao, L. Wang, Enhanced radiation-tolerant oxide dispersion strengthened steel and its microstructure evolution under helium-implantation and heavy-ion Irradiation, Sci. Rep. 7 (2017) 40343.
[19] G.R. Odette, Recent progress in developing and qualifying nanostructured ferritic alloys for advanced fission and fusion applications, JOM 66(12) (2014) 2427-2441.

[20] X.M. Bai, A.F. Voter, R.G. Hoagland, M. Nastasi, B.P. Uberuaga, Efficient annealing of radiation damage near grain boundaries via interstitial emission, Science 327(5973) (2010) 1631-1634.
[21] W.Z. Han, M.J. Demkowicz, E.G. Fu, Y.Q. Wang, A. Misra, Effect of grain boundary character on sink efficiency, Acta Mater. 60(18) (2012) 6341-6351.
[22] N. Nita, R. Schaeublin, M. Victoria, Impact of irradiation on the microstructure of nanocrystalline materials, J. Nucl. Mater. 329 (2004) 953-957.
[23] Y. Chimi, A. Iwase, N. Ishikawa, M. Kobiyama, T. Inami, S. Okuda, Accumulation and recovery of defects in ion-irradiated nanocrystalline gold, J. Nucl. Mater. 297(3) (2001) 355-357.
[24] M. Rose, A.G. Balogh, H. Hahn, Instability of irradiation induced defects in nanostructured materials, Nucl. Instrum. Methods Phys. Res., Sect. B 127 (1997) 119-122.
[25] G. Palumbo, S.J. Thorpe, K.T. Aust, On the contribution of triple junctions to the structure and properties of nanocrystalline materials, Scripta Metall. et Mater. 24(7) (1990) 1347-1350.
[26] C.M. Barr, N. Li, B.L. Boyce, K. Hattar, Examining the influence of grain size on radiation tolerance in the nanocrystalline regime, Appl. Phys. Lett. 112(18) (2018) 181903.
[27] O. El-Atwani, J.E. Nathaniel, A.C. Leff, B.R. Muntifering, J.K. Baldwin, K. Hattar, M.L. Taheri, The role of grain size in He bubble formation: implications for swelling resistance, J. Nucl. Mater. 484 (2017) 236-244.
[28] C.M. Barr, O. El-Atwani, D. Kaoumi, K. Hattar, Interplay between grain boundaries and radiation damage, JOM (2019) 1-12.
[29] P.R. Cantwell, M. Tang, S.J. Dillon, J. Luo, G.S. Rohrer, M.P. Harmer, Grain boundary complexions, Acta Mater. 62 (2014) 1-48.
[30] S.J. Dillon, M. Tang, W.C. Carter, M.P. Harmer, Complexion: a new concept for kinetic engineering in materials science, Acta Mater. 55(18) (2007) 6208-6218.
[31] J. Luo, H. Cheng, K.M. Asl, C.J. Kiely, M.P. Harmer, The role of a bilayer interfacial phase on liquid metal embrittlement, Science 333(6050) (2011) 1730-1733.
[32] J. Luo, V.K. Gupta, D.H. Yoon, H.M. Meyer III, Segregation-induced grain boundary premelting in nickel-doped tungsten, Appl. Phys. Lett. 87(23) (2005) 231902.
[33] J. Luo, Liquid-like interface complexion: From activated sintering to grain boundary diagrams, Curr. Opin. Solid State Mater. Sci. 12(5) (2008) 81-88.
[34] A. Khalajhedayati, T.J. Rupert, High-Temperature Stability and Grain Boundary Complexion Formation in a Nanocrystalline Cu-Zr Alloy, JOM 67(12) (2015) 2788-2801.
[35] S.J. Fensin, D. Olmsted, D. Buta, M. Asta, A. Karma, J.J. Hoyt, Structural disjoining potential for grain-boundary premelting and grain coalescence from molecular-dynamics simulations, Physical Review E 81(3) (2010) 031601.
[36] Z. Pan, T.J. Rupert, Effect of grain boundary character on segregation-induced structural transitions, Phys. Rev. B 93(13) (2016) 134113.
[37] Y. Mishin, W.J. Boettinger, J.A. Warren, G.B. McFadden, Thermodynamics of grain boundary premelting in alloys. I. Phase-field modeling, Acta Mater. 57(13) (2009) 3771-3785.
[38] P.L. Williams, Y. Mishin, Thermodynamics of grain boundary premelting in alloys. II. Atomistic simulation, Acta Mater. 57(13) (2009) 3786-3794.
[39] C.M. Grigorian, T.J. Rupert, Thick amorphous complexion formation and extreme thermal stability in ternary nanocrystalline Cu-Zr-Hf alloys, Acta Mater. 179 (2019) 172-182.
[40] A. Khalajhedayati, Z. Pan, T.J. Rupert, Manipulating the interfacial structure of nanomaterials to achieve a unique combination of strength and ductility, Nat. Comm. 7 (2016) 10802.

[41] X. Shi, J. Luo, Grain boundary wetting and prewetting in Ni-doped Mo, Appl. Phys. Lett. 94(25) (2009) 251908.
[42] J.D. Schuler, O.K. Donaldson, T.J. Rupert, Amorphous complexions enable a new region of high temperature stability in nanocrystalline Ni-W, Scripta Mater. 154 (2018) 49-53.
[43] G. Odette, M. Alinger, B. Wirth, Recent developments in irradiation-resistant steels, Annu. Rev. Mater. Res. 38 (2008) 471-503.
[44] S. Ukai, M. Fujiwara, Perspective of ODS alloys application in nuclear environments, J. Nucl. Mater. 307 (2002) 749-757.
[45] T. Chookajorn, H.A. Murdoch, C.A. Schuh, Design of stable nanocrystalline alloys, Science 337 (2012) 951-954.
[46] E. Nes, N. Ryum, O. Hunderi, On the Zener drag, Acta Metall. 33(1) (1985) 11-22.
[47] K.A. Darling, A.J. Roberts, Y. Mishin, S.N. Mathaudhu, L.J. Kecskes, Grain size stabilization of nanocrystalline copper at high temperatures by alloying with tantalum, J. Alloys Compd. 573 (2013) 142-150.
[48] A.J. Detor, C.A. Schuh, Microstructural evolution during the heat treatment of nanocrystalline alloys, J. Mater. Res. 22(11) (2007) 3233-3248.
[49] Z.B. Jiao, C.A. Schuh, Nanocrystalline Ag-W alloys lose stability upon solute desegregation from grain boundaries, Acta Mater. 161 (2018) 194-206.
[50] E. Botcharova, J. Freudenberger, L. Schultz, Cu–Nb alloys prepared by mechanical alloying and subsequent heat treatment, J. Alloys Compd. 365(1-2) (2004) 157-163.
[51] J.E. Ludy, T.J. Rupert, Amorphous intergranular films act as ultra-efficient point defect sinks during collision cascades, Scripta Mater. 110 (2016) 37-40.
[52] X. Zhang, N. Li, O. Anderoglu, H. Wang, J.G. Swadener, T. Höchbauer, A. Misra, R.G. Hoagland, Nanostructured Cu/Nb multilayers subjected to helium ion-irradiation, Nucl. Instrum. Methods Phys. Res., Sect. B 261(1-2) (2007) 1129-1132.
[53] E.G. Fu, J. Carter, G. Swadener, A. Misra, L. Shao, H. Wang, X. Zhang, Size dependent enhancement of helium ion irradiation tolerance in sputtered Cu/V nanolaminates, J. Nucl. Mater. 385(3) (2009) 629-632.
[54] T. Höchbauer, A. Misra, K. Hattar, R. Hoagland, Influence of interfaces on the storage of ion-implanted He in multilayered metallic composites, J. Appl. Phys. 98(12) (2005) 123516.
[55] M. Zhernenkov, S. Gill, V. Stanic, E. DiMasi, K. Kisslinger, J.K. Baldwin, A. Misra, M.J. Demkowicz, L. Ecker, Design of radiation resistant metallic multilayers for advanced nuclear systems, Appl. Phys. Lett. 104(24) (2014) 241906.
[56] M.J. Demkowicz, D. Bhattacharyya, I. Usov, Y.Q. Wang, M. Nastasi, A. Misra, The effect of excess atomic volume on He bubble formation at fcc–bcc interfaces, Appl. Phys. Lett. 97(16) (2010) 161903.
[57] A. Kashinath, A. Misra, M.J. Demkowicz, Stable storage of helium in nanoscale platelets at semicoherent interfaces, Phys. Rev. Lett. 110(8) (2013) 086101.
[58] M. Samaras, P.M. Derlet, H. Van Swygenhoven, M. Victoria, Radiation damage near grain boundaries, Philos. Mag. 83(31-34) (2003) 3599-3607.
[59] Q. Su, L. Price, J.A. Colon Santana, L. Shao, M. Nastasi, Irradiation tolerance of amorphous SiOC/crystalline Fe composite, Mater. Lett. 155 (2015) 138-141.
[60] J. Luo, Developing interfacial phase diagrams for applications in activated sintering and beyond: current status and future directions, J. Am. Ceram. Soc. 95(8) (2012) 2358-2371.
[61] B.P. Uberuaga, L.J. Vernon, E. Martinez, A.F. Voter, The relationship between grain boundary structure, defect mobility, and grain boundary sink efficiency, Sci. Rep. 5 (2015) 9095.

[62] J.D. Schuler, T.J. Rupert, Materials selection rules for amorphous complexion formation in binary metallic alloys, Acta Mater. 140 (2017) 196-205.
[63] X. Shi, J. Luo, Developing grain boundary diagrams as a materials science tool: a case study of nickel-doped molybdenum, Phys. Rev. B 84(1) (2011).
[64] Z. Pan, T.J. Rupert, Effect of grain boundary character on segregation-induced structural transitions, Physical Review B 93(13) (2016) 134113.
[65] P.C. Millett, R.P. Selvam, A. Saxena, Stabilizing nanocrystalline materials with dopants, Acta Mater. 55(7) (2007) 2329-2336.
[66] H. Fecht, E. Hellstern, Z. Fu, W. Johnson, Nanocrystalline metals prepared by high-energy ball milling, Metall. Trans. A 21(9) (1990) 2333.
[67] K. Hattar, D.C. Bufford, D.L. Buller, Concurrent in situ ion irradiation transmission electron microscope, Nucl. Instrum. Methods Phys. Res., Sect. B 338 (2014) 56-65.
[68] J.F. Ziegler, M.D. Ziegler, J.P. Biersack, SRIM–The stopping and range of ions in matter (2010), Nucl. Instrum. Methods Phys. Res., Sect. B 268(11-12) (2010) 1818-1823.
[69] R.E. Stoller, M.B. Toloczko, G.S. Was, A.G. Certain, S. Dwaraknath, F.A. Garner, On the use of SRIM for computing radiation damage exposure, Nucl. Instrum. Methods Phys. Res., Sect. B 310 (2013) 75-80.
[70] D.B. Bober, A. Khalajhedayati, M. Kumar, T.J. Rupert, Grain boundary character distributions in nanocrystalline metals produced by different processing routes, Metall. Mater. Trans. A 47(3) (2016) 1389-1403.
[71] R.R. Keller, R.H. Geiss, Transmission EBSD from 10 nm domains in a scanning electron microscope, J. Microscopy 245(3) (2012) 245-251.
[72] P.W. Trimby, Orientation mapping of nanostructured materials using transmission Kikuchi diffraction in the scanning electron microscope, Ultramicroscopy 120 (2012) 16-24.
[73] I.S. Kim, J.D. Hunn, N. Hashimoto, D.L. Larson, P.J. Maziasz, K. Miyahara, E.H. Lee, Defect and void evolution in oxide dispersion strengthened ferritic steels under 3.2 MeV Fe+ ion irradiation with simultaneous helium injection, J. Nucl. Mater. 280(3) (2000) 264-274.
[74] O. El-Atwani, E. Martinez, E. Esquivel, M. Efe, C. Taylor, Y.Q. Wang, B.P. Uberuaga, S.A. Maloy, Does sink efficiency unequivocally characterize how grain boundaries impact radiation damage?, Phys. Rev. Mater. 2(11) (2018) 113604.
[75] P. Thevenaz, U.E. Ruttimann, M. Unser, A pyramid approach to subpixel registration based on intensity, IEEE Transactions on Image Processing 7(1) (1998) 27-41.
[76] M.L. Jenkins, M.A. Kirk, Characterisation of Radiation Damage by Transmission Electron Microscopy, CRC Press, 2000.
[77] S.J. Zinkle, G.L. Kulcinski, R.W. Knoll, Microstructure of copper following high dose 14 MeV Cu ion irradiation, J. Nucl. Mater. 138(1) (1986) 46-56.
[78] N. Yoshida, K. Kitajima, E. Kuramoto, Evolution of damage structures under 14 MeV neutron, 4 MeV Ni ion and 1.25 MeV electron irradiation, J. Nucl. Mater. 122(1-3) (1984) 664-668.
[79] A.J.E. Foreman, M.J. Makin, Dislocation movement through random arrays of obstacles, Philos. Mag. 14(131) (1966) 911-924.
[80] J.S. Robach, I.M. Robertson, B.D. Wirth, A. Arsenlis, In-situ transmission electron microscopy observations and molecular dynamics simulations of dislocation-defect interactions in ion-irradiated copper, Philos. Mag. 83(8) (2003) 955-967.

[81] R.C. Hugo, H. Kung, J.R. Weertman, R. Mitra, J.A. Knapp, D.M. Follstaedt, In-situ TEM tensile testing of DC magnetron sputtered and pulsed laser deposited Ni thin films, Acta Mater. 51(7) (2003) 1937-1943.
[82] O. El-Atwani, J.E. Nathaniel, A.C. Leff, K. Hattar, M.L. Taheri, Direct observation of sink-dependent defect evolution in nanocrystalline iron under irradiation, Sci. Rep. 7(1) (2017) 1836.
[83] W. Jäger, M. Rühle, M. Wilkens, Elastic interaction of a dislocation loop with a traction- free surface, Physica Status Solidi A 31(2) (1975) 525-533.
[84] W. Jäger, M. Wilkens, Formation of vacancy- type dislocation loops in tungsten bombarded by 60 keV Au ions, Physica Status Solidi A 32(1) (1975) 89-100.
[85] B. Muntifering, R. Dingreville, K. Hattar, J. Qu, Electron beam effects during in-situ annealing of self-ion irradiated nanocrystalline nickel, MRS Online Proc. Library Archive 1809 (2015) 13-18.
[86] C.W. Allen, In situ ion- and electron-irradiation effects studies in transmission electron microscopes, Ultramicroscopy 56(1) (1994) 200-210.
[87] J.A. Sundararajan, M. Kaur, Y. Qiang, Mechanism of electron beam induced oxide layer thickening on iron–iron oxide core–shell nanoparticles, J. Phys. Chem. C 119(15) (2015) 8357-8363.
[88] V. Randle, O. Engler, Introduction to Texture Analysis: Macrotexture, Microtexture and Orientation Mapping, CRC press, 2014.
[89] S.J. Zinkle, Radiation-Induced Effects on Microstructure, Comprehensive Nuclear Materials, Elsevier, Amsterdam, 2012, pp. 65-98.
[90] J. Luo, X. Shi, Grain boundary disordering in binary alloys, Appl. Phys. Lett. 92(10) (2008) 101901.
[91] M.A. Kirk, M. Li, D. Xu, B.D. Wirth, Predicting neutron damage using TEM with in situ ion irradiation and computer modeling, J. Nucl. Mater. 498 (2018) 199-212.
[92] J.R. Michael, Focused ion beam induced microstructural alterations: texture development, grain growth, and intermetallic formation, Microsc. Microanal. 17(3) (2011) 386-397.
[93] Z. Pan, T.J. Rupert, Amorphous intergranular films as toughening structural features, Acta Mater. 89 (2015) 205-214.
[94] D. Kaoumi, A.T. Motta, R.C. Birtcher, A thermal spike model of grain growth under irradiation, J. Appl. Phys. 104(7) (2008) 073525.
[95] D.C. Bufford, F.F. Abdeljawad, S.M. Foiles, K. Hattar, Unraveling irradiation induced grain growth with in situ transmission electron microscopy and coordinated modeling, Appl. Phys. Lett. 107(19) (2015) 191901.
[96] M. Alurralde, A. Caro, M. Victoria, Radiation damage cascades: liquid droplet treatment of subcascade interactions, J. Nucl. Mater. 183(1-2) (1991) 33-45.
[97] J.C. Liu, M. Nastasi, J.W. Mayer, Ion irradiation induced grain growth in Pd polycrystalline thin films, J. Appl. Phys. 62(2) (1987) 423-428.
[98] E.A. Holm, S.M. Foiles, How grain growth stops: a mechanism for grain-growth stagnation in pure materials, Science 328(5982) (2010) 1138-1141.
[99] C.J. O'Brien, C.M. Barr, P.M. Price, K. Hattar, S.M. Foiles, Grain boundary phase transformations in PtAu and relevance to thermal stabilization of bulk nanocrystalline metals, J. Mater. Sci. 53(4) (2018) 2911-2927.
[100] C.M. Barr, G.A. Vetterick, K.A. Unocic, K. Hattar, X.M. Bai, M.L. Taheri, Anisotropic radiation-induced segregation in 316L austenitic stainless steel with grain boundary character, Acta Mater. 67 (2014) 145-155.

[101] T.R. Allen, J.T. Busby, G.S. Was, E.A. Kenik, On the mechanism of radiation-induced segregation in austenitic Fe–Cr–Ni alloys, J. Nucl. Mater. 255(1) (1998) 44-58.
[102] H. Wiedersich, P.R. Okamoto, N.Q. Lam, A theory of radiation-induced segregation in concentrated alloys, J. Nucl. Mater. 83(1) (1979) 98-108.
[103] G.S. Was, J.P. Wharry, B. Frisbie, B.D. Wirth, D. Morgan, J.D. Tucker, T.R. Allen, Assessment of radiation-induced segregation mechanisms in austenitic and ferritic–martensitic alloys, J. Nucl. Mater. 411(1) (2011) 41-50.
[104] R.S. Averback, Fundamental aspects of ion beam mixing, Nucl. Instrum. Methods Phys. Res., Sect. B 15(1) (1986) 675-687.
[105] N. Li, M.S. Martin, O. Anderoglu, A. Misra, L. Shao, H. Wang, X. Zhang, He ion irradiation damage in A1∕Nb multilayers, J. Appl. Phys. 105(12) (2009) 123522.
[106] N. Li, E.G. Fu, H. Wang, J.J. Carter, L. Shao, S.A. Maloy, A. Misra, X. Zhang, He ion irradiation damage in Fe/W nanolayer films, J. Nucl. Mater. 389(2) (2009) 233-238.
[107] Q. Su, B. Cui, M.A. Kirk, M. Nastasi, In-situ observation of radiation damage in nano-structured amorphous SiOC/crystalline Fe composite, Scripta Mater. 113 (2016) 79-83.

| Experiment Type | Temperature (°C) | Ion Beam | Power (MeV) | Flux (ions/cm²s) | Fluence (ions/cm²) | Proj. Ion Beam Penetration Depth (µm) | Max Dose (dpa) | Max Implanted Au (at.%) |
|---|---|---|---|---|---|---|---|---|
| In situ TEM | 25 | Au$^{4+}$ | 2.8 | $1.5 \times 10^{11}$ | $4.0 \times 10^{14}$ | NA | 6.8 | 0.03 |
| Ex situ | 25 | Au$^{4+}$ | 20 | $1.1 \times 10^{12}$ | $2.1 \times 10^{16}$ | 2.2 +/-0.3 | 95 | 0.3 |
| Ex situ | 200 | Au$^{4+}$ | 20 | $1.1 \times 10^{12}$ | $2.1 \times 10^{16}$ | 2.2 +/-0.3 | 95 | 0.3 |

**Table 1**. Details of the in situ TEM and ex situ irradiation experiments.

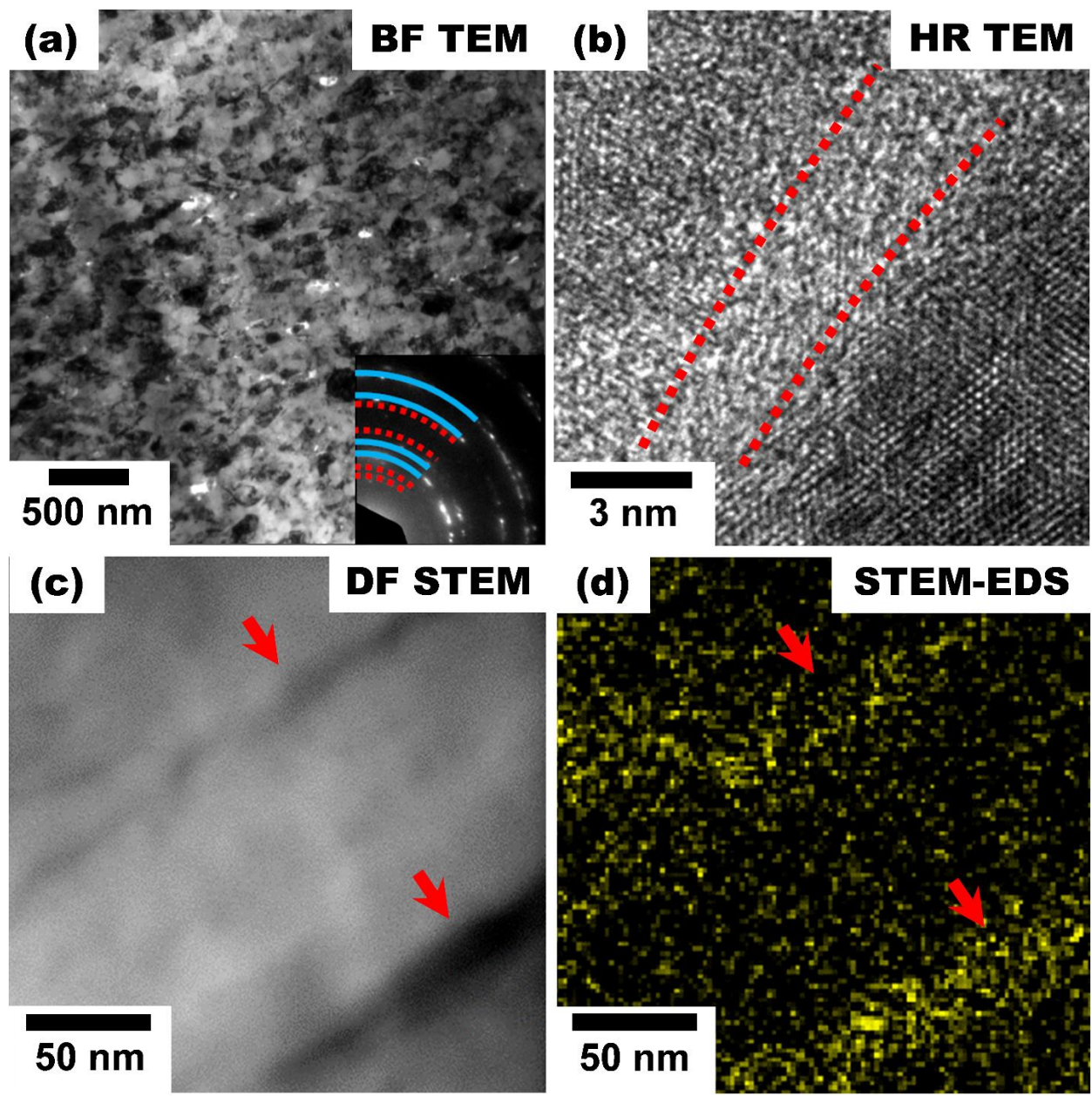

**Figure 1**. Microstructural and chemical analysis of the Cu-3 at.% Zr alloy that was heat treated but before irradiation. (a) Bright field (BF) TEM image of the grain structure in the sample containing AIFs. The inset shows the associated selected area electron diffraction pattern, where the solid blue lines indicate the face centered cubic Cu phase and the dashed red lines indicate the ZrC phase. (b) High resolution (HR) TEM image of an AIF, where the dashed red lines highlight the presence of the AIF along the grain boundary. (c) Dark field (DF) STEM image and (d) the associated EDS map, where Zr signals are indicated with yellow. The red arrows in (c) and (d) indicate grain boundaries with Zr segregation.

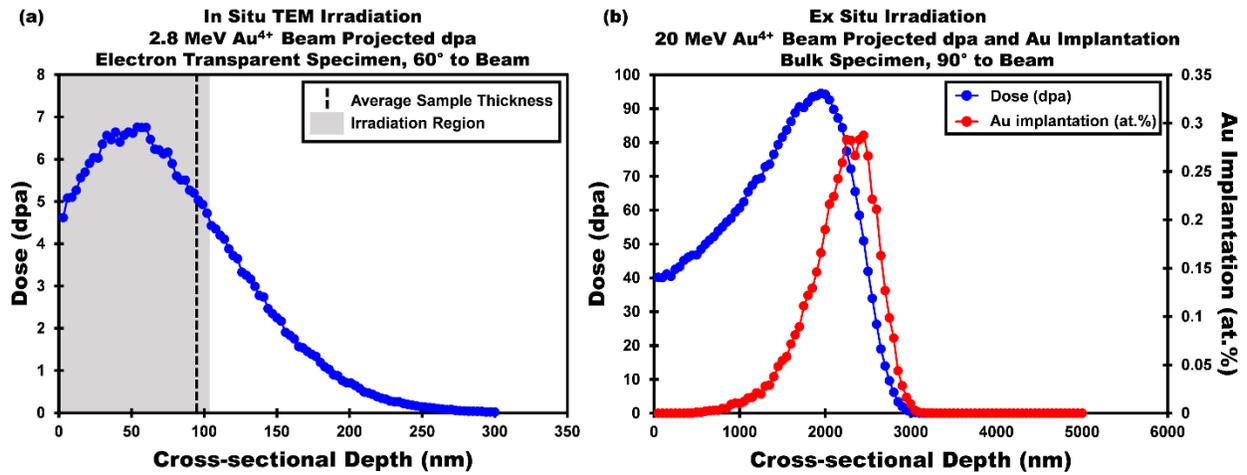

**Figure 2**. The estimated damage accumulation levels measured in dpa as a function of depth into the sample for the in situ TEM and ex situ irradiation experiments. (a) In situ TEM irradiation profile using a 2.8 MeV Au$^{4+}$ beam incident 60° to the sample surface, where dose is represented by blue circles. The dashed black line at 95 nm indicates the average thickness of the TEM samples and the gray box indicates the approximate irradiated region of the TEM foil. (b) Ex situ irradiation profile using a 20 MeV Au$^{4+}$ beam incident 90° to the sample surface. Dose is represented by blue circles, and the amount of implanted Au by red circles.

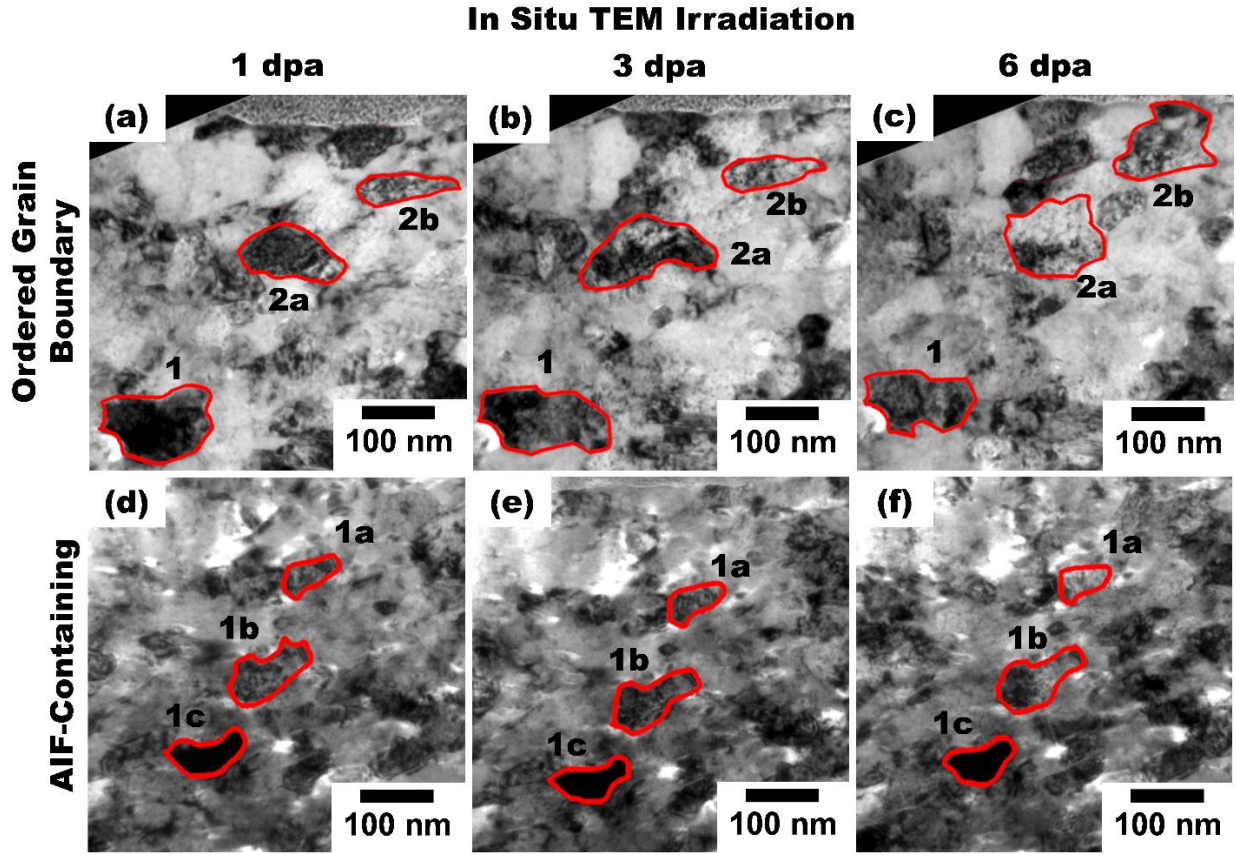

**Figure 3**. Microstructural evolution with increasing dose measured in dpa during in situ TEM irradiation. (a)-(c) Microstructural evolution of the ordered grain boundary sample with increasing dose at 1, 3, and 6 dpa, respectively. (d)-(f) Microstructural evolution of the sample containing AIFs with increasing dose at 1, 3, and 6 dpa, respectively. Two types of microstructural evolution were observed. Type 1 refers to a change in grain shape without a significant change in area. Type 2 refers to grain growth indicated by a significant increase in grain area. Grains indicative of each microstructural evolution mode are highlighted in red and numbered in accordance with their evolution mode, with lettering indicating multiple grans that experienced an evolution type.

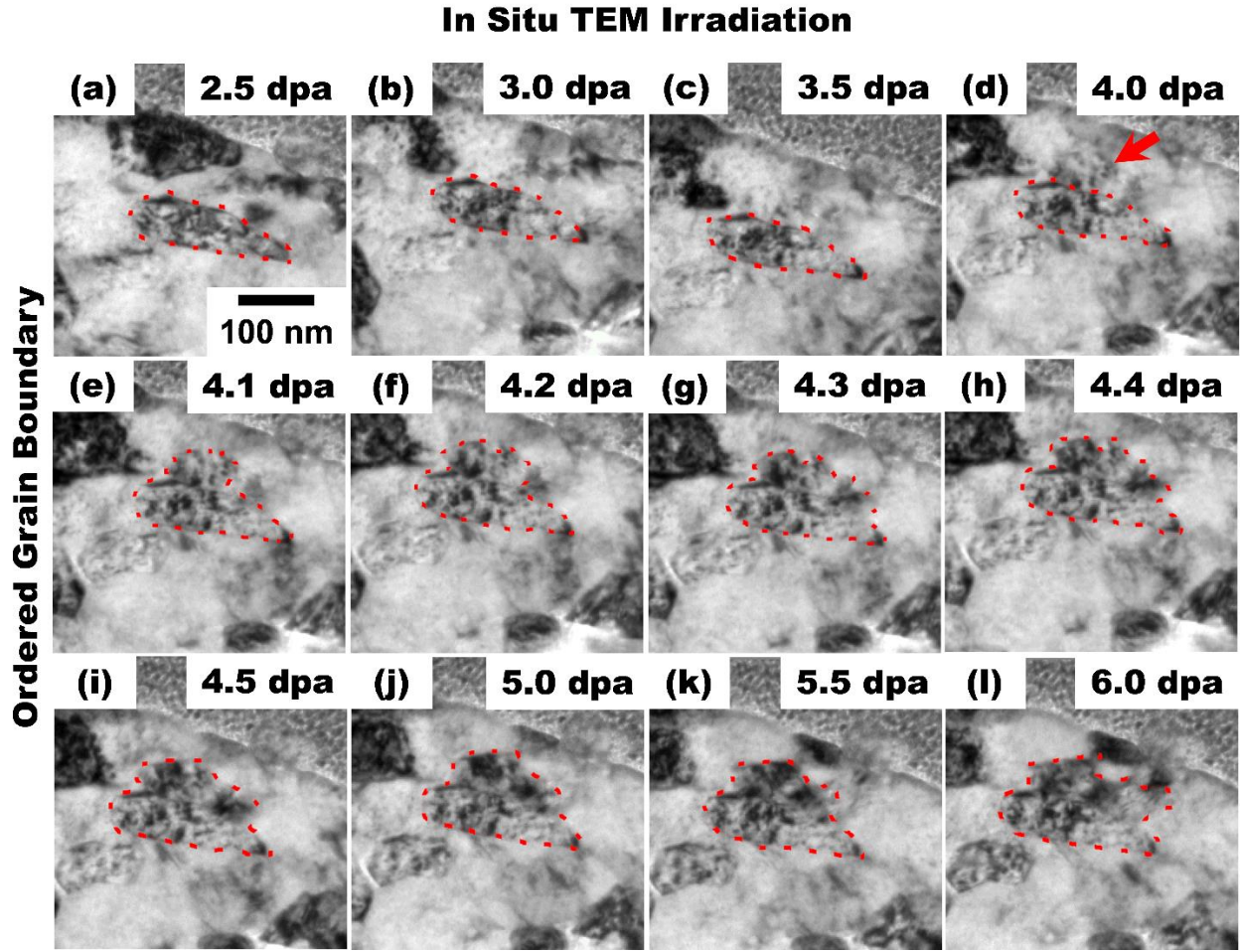

**Figure 4**. Greater detail of the microstructural evolution and grain growth of grain 2b from the ordered grain boundary sample presented in Figure 3. The red arrow in (d) indicates a region of significant contrast change where the grain then grows.

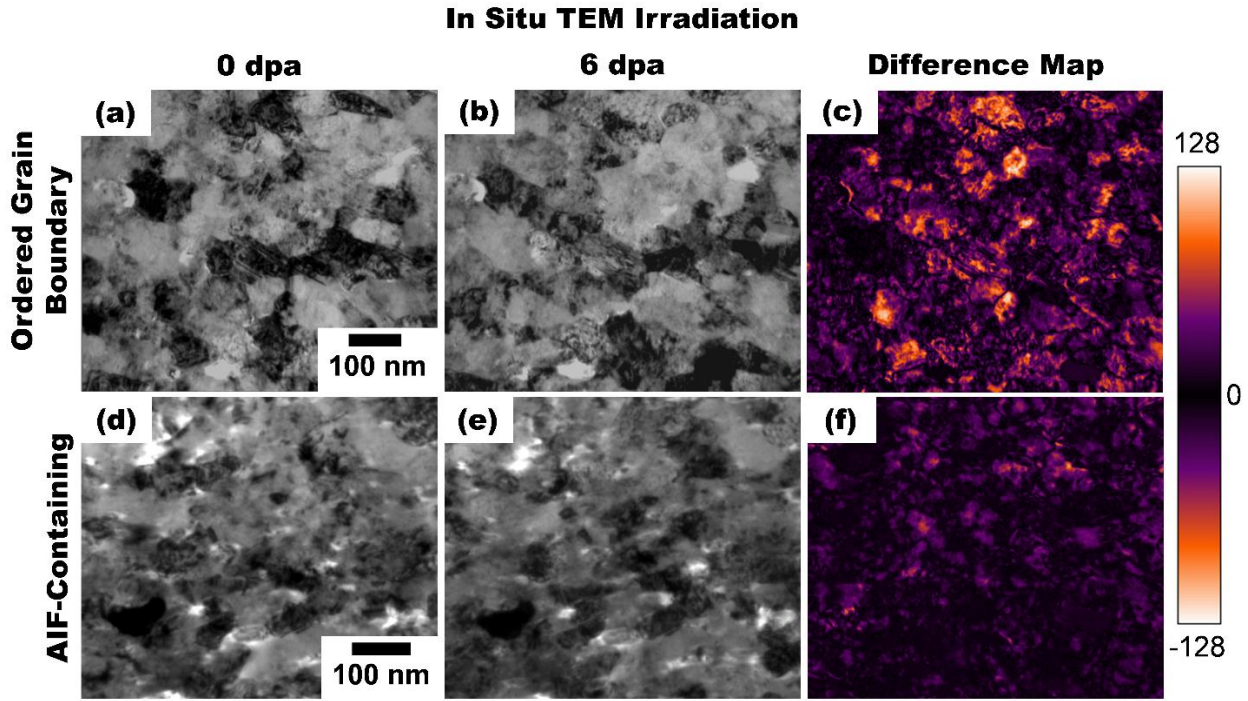

**Figure 5**. Difference maps generated by finding contrast shifts that occur between the beginning and end of irradiation (0 and 6 dpa, respectively) for each sample. (a) and (b) show the bright field TEM images (normalized brightness and contrast) for the ordered grain boundary sample, and (d) and (e) show similar images for the AIF-containing sample. (c) and (f) show the difference maps, where orange and white indicate a large contrast shift in the microstructure while black and purple indicate a small shift.

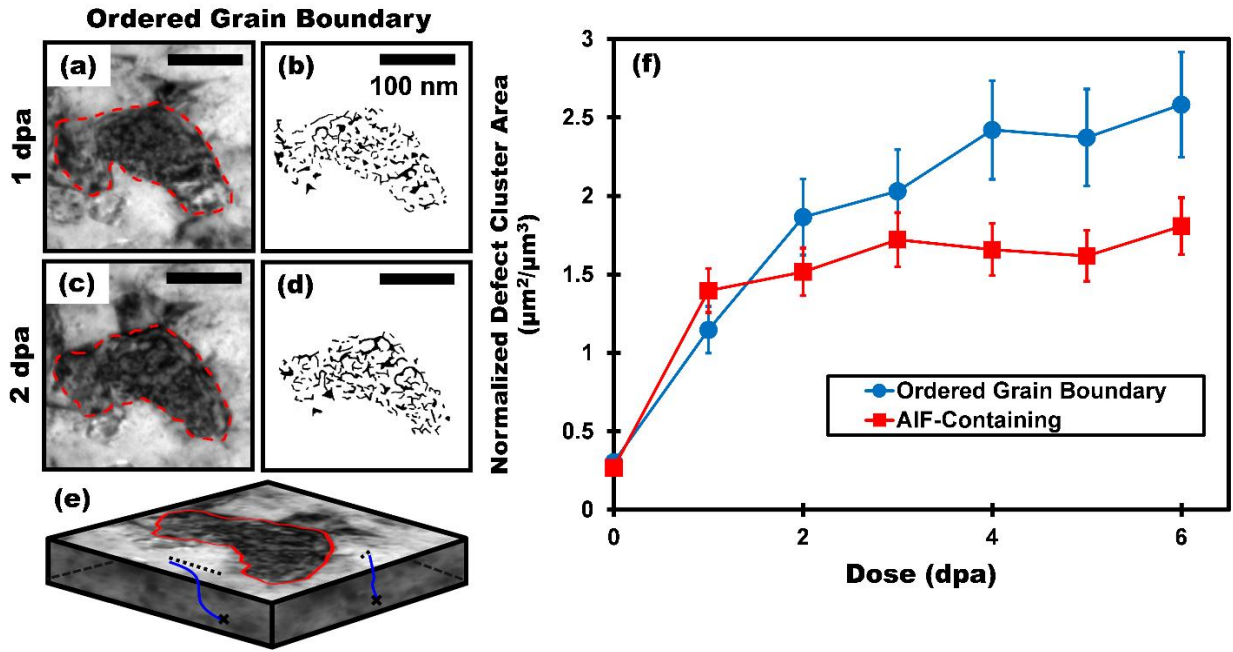

**Figure 6**. The normalized defect cluster area density with increasing dose from 0 to 6 dpa. The scale bars in (a)-(d) are 100 nm. (a) and (c) show bright field TEM images of a representative grain from the ordered grain boundary sample subjected to in situ TEM irradiation at 1 and 2 dpa, with the grain boundary outlined in red. (b) and (d) show radiation induced defects, marked in black, that are present in each grain. (e) shows the same grain from (a) and (c) rotated to schematically show the impact of specimen thickness on the measurement of defect cluster area. (f) Normalized defect cluster area per volume with increasing dose for the ordered grain boundary (blue circles) and AIF-containing (red squares) samples. Error bars represent the upper and lower bounds of defect density due to variations in sample thickness.

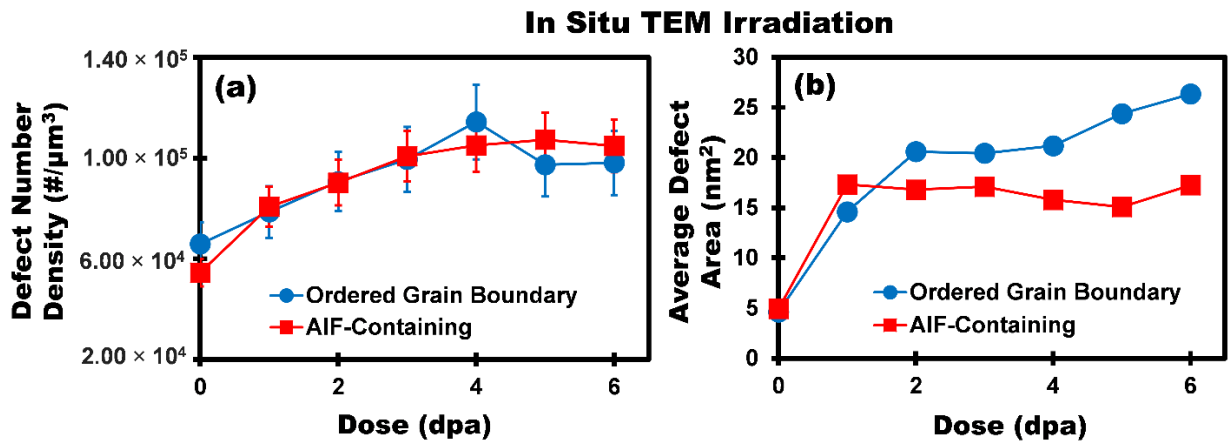

**Figure 7**. (a) The defect number density and (b) average defect area for the ordered grain boundary (blue circles) and AIF-containing (red squares) samples with increasing dose during in situ TEM irradiation experiments. Error bars represent the upper and lower bounds of defect density due to variations in sample thickness.

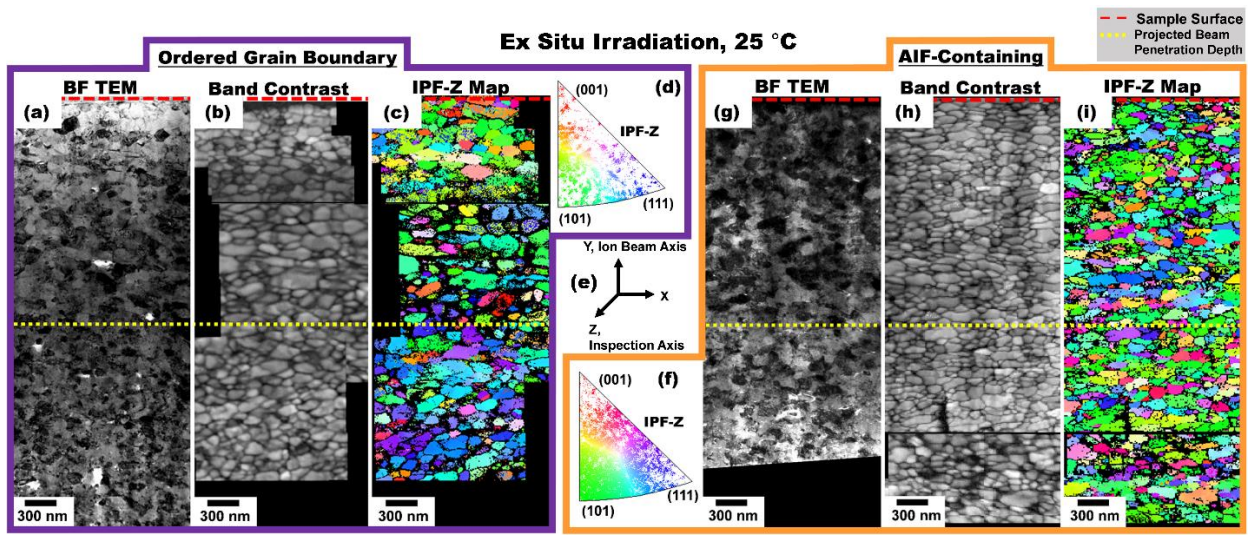

**Figure 8**. (a)-(c) Bright field TEM, Kikuchi band contrast, and orientation maps for the ordered grain boundary sample after ex situ irradiation at 25 °C (outlined in purple). (g)-(i) Bright field TEM, Kikuchi band contrast, and orientation maps for the sample containing AIFs (outlined in orange). The dashed red lines show the sample surface and the dashed yellow lines show the projected ion beam penetration depth. (e) The sample reference frame axes, where the Y-axis is parallel to the ion beam axis and the Z-axis is parallel to the inspection axis. (d) and (f) Equal area projection plots for the corresponding IPF-Z maps, where (d) refers to the ordered grain boundary sample and (f) refers to the AIF-containing sample.

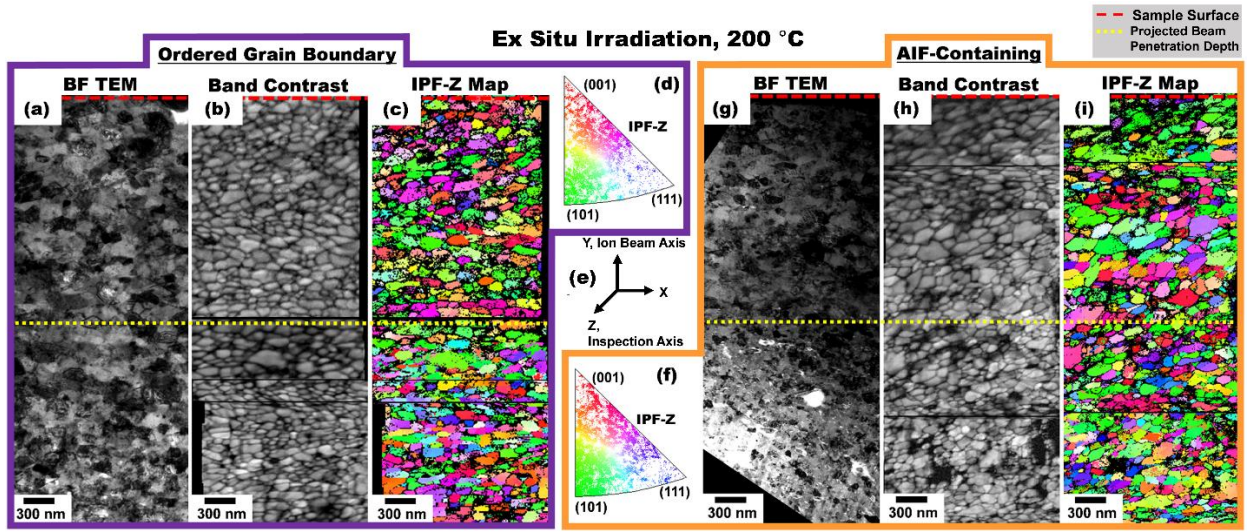

**Figure 9**. (a)-(c) Bright field TEM, Kikuchi band contrast, and orientation maps for the ordered grain boundary sample after ex situ irradiation at 200 °C (outlined in purple). (g)-(i) Bright field TEM, Kikuchi band contrast, and orientation maps for the sample containing AIFs (outlined in orange). The dashed red lines show the sample surface and the dashed yellow lines show the projected ion beam penetration depth. (e) The sample reference frame axes, where the Y-axis is parallel to the ion beam axis and the Z-axis is parallel to the inspection axis. (d) and (f) Equal area projection plots for the corresponding IPF-Z maps, where (d) refers to the ordered grain boundary sample and (f) refers to the AIF-containing sample.

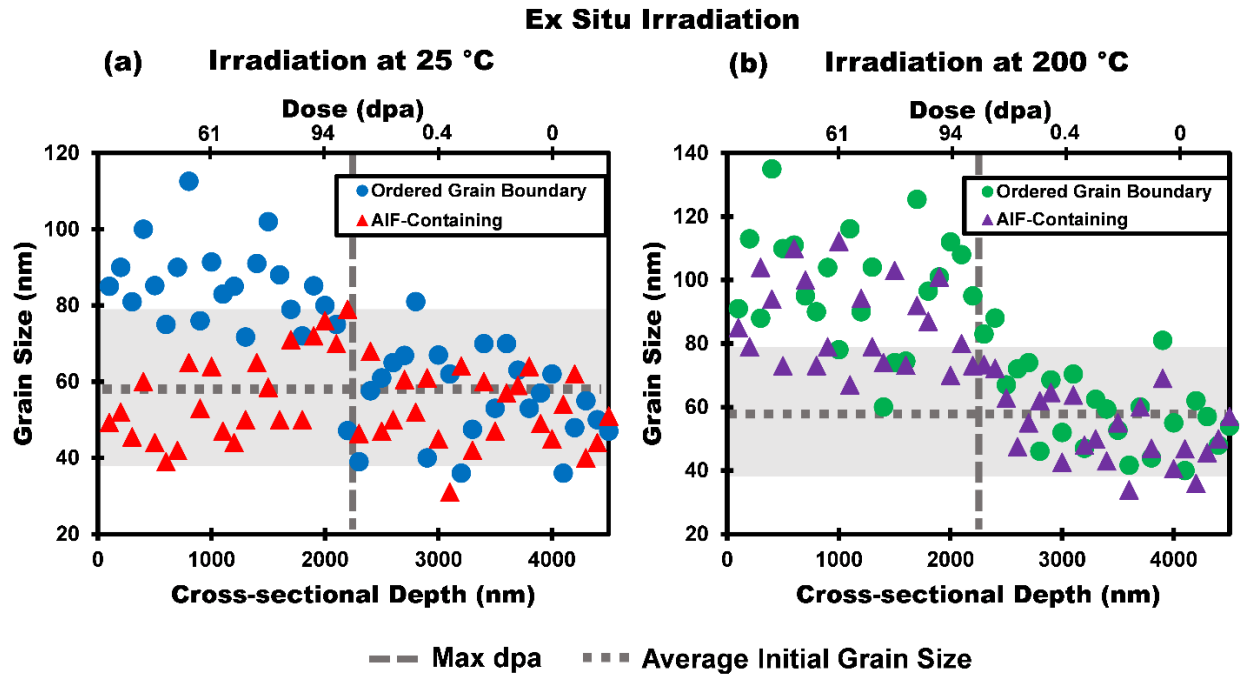

**Figure 10**. Grain size as a function of depth for the ex situ irradiated samples. Each point shows the average grain size within a 100 nm depth segment. The vertical dashed gray lines show the region of maximum dose which coincides with the projection ion beam penetration depth at approximately 2.2 μm. The horizontal dashed gray lines show the average grain size before irradiation and the shaded box shows the standard deviation. (a) Grain size measurements for the ordered grain boundary (blue circles) and the AIF-containing (red triangles) samples irradiated at 25 °C. (b) Grain size measurements for the ordered grain boundary (green circles) and AIF-containing (purple triangles) samples irradiated at 200 °C.

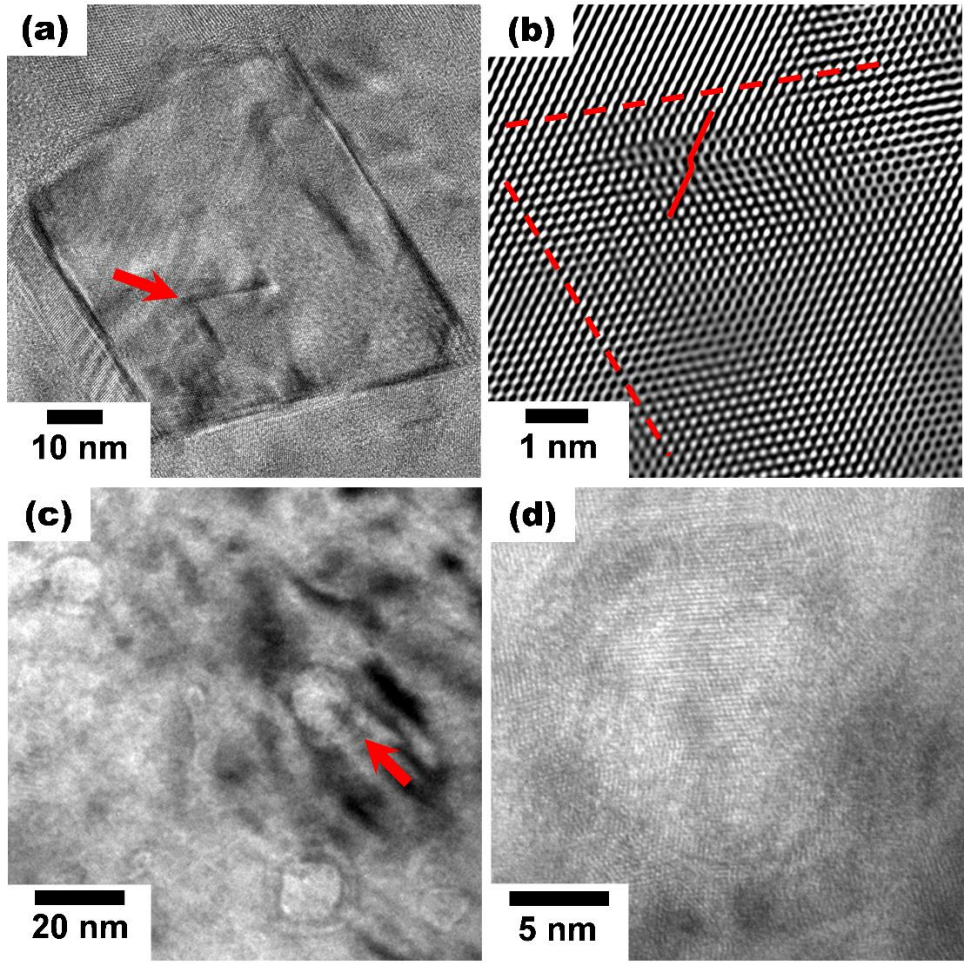

**Figure 11**. TEM images of defects observed in the ex situ irradiated samples. (a) Several defects appear as black spot damage which may be point defect clusters, stacking fault tetrahedra, loops, dislocations, or dislocation tangles. (b) Inverse fast Fourier transform filtered image of the possible stacking fault tetrahedron that was indicated by the red arrow in (a). Dotted red lines frame the stacking fault tetrahedron and the solid red line indicates a possible stacking fault. (c) Several under-focused cavities, with one cavity indicated by a red arrow and shown in greater detail in (d).

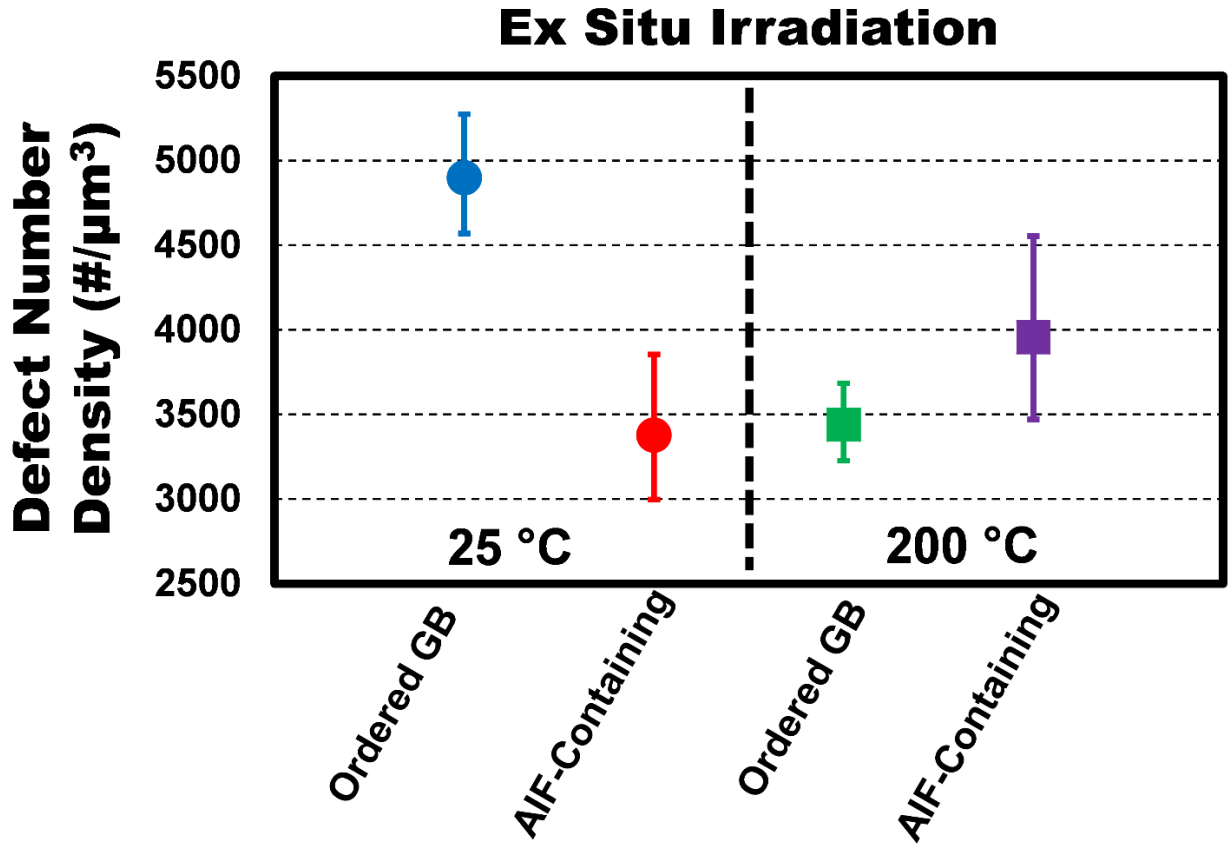

**Figure 12.** The defect number density for the ordered grain boundary and AIF-containing samples after ex situ irradiation at 25 °C and 200 °C. For the samples irradiated at 25 °C, the ordered grain boundary sample is shown with a blue circle and the AIF-containing sample with a red circle. For the 200 °C irradiation, the ordered grain boundary sample is shown with a green square and the AIF-containing sample with a purple square. The associated ranges indicate the upper and lower error bounds of defect density due to variations in sample thickness.